%% file: Paper.tex
\documentclass[11pt,letterpaper]{article}

\usepackage{mathrsfs}           
\usepackage{vmargin}   
\usepackage{algorithm}
\usepackage{algorithmic}
\usepackage{tikz}
\usepackage{bm}
\usepackage{enumerate}

\usepackage{amsmath,amssymb}    
\usepackage{verbatim}           
\usepackage{xspace}             
\usepackage{graphicx,float}     
\usepackage{ifthen,calc}        
\usepackage{textcomp}           
\usepackage{fancybox}           
\usepackage{hhline}             
\usepackage{float}              

\usepackage[rflt]{floatflt}     
\usepackage[small,compact]{titlesec}
\usepackage{setspace}
\usepackage{subfigure}

\usepackage{color}
\definecolor{ForestGreen}{rgb}{0.1333,0.5451,0.1333}
\usepackage[dvips]{thumbpdf}

\setpapersize{USletter}
\setmarginsrb{.75in}{.5in}        
             {.75in}{.5in}        
             {.25in}{.25in}     
             {.25in}{.5in}      
\newcommand{\showccc}[0]{0}
\newcommand{\ccc}[2][nothing]{
  \ifthenelse{\showccc=0}{}{
    \ensuremath{^{\Lsh\Rsh}}\marginpar{\raggedright\tiny\textsf{%
        \ifthenelse{\equal{#1}{nothing}}{}{\textbf{#1}\\}#2}}}}

\newcounter{hours}\newcounter{minutes}
\newcommand{\hhmm}{%
  \setcounter{hours}{\time/60}%
  \setcounter{minutes}{\time-\value{hours}*60}%
  \ifthenelse{\value{hours}<10}{0}{}\thehours:%
  \ifthenelse{\value{minutes}<10}{0}{}\theminutes}

\newtheorem{theorem}{Theorem}[section]

\newtheorem{corollary}[theorem]{Corollary}
\newtheorem{definition}[theorem]{Definition}

\newtheorem{lemma}[theorem]{Lemma}
\newtheorem{fact}[theorem]{Fact}

\newcommand{\Proof}[0]{\smallskip\noindent\textit{\textbf{Proof}}\quad}
\newcommand{\Proofof}[1]{\smallskip\noindent\textit{\textbf{Proof of #1:}}\quad}

\newcommand{\QED}[0]{\hfill\ensuremath{\blacksquare}\medspace\vskip 2ex}

\floatstyle{ruled}
\newfloat{algo}{tbp}{lop}
\floatname{algo}{Algorithm}

\usepackage{float}
\floatstyle{plain}\newfloat{myfig}{t}{figs}[section]
\floatname{myfig}{\textsc{Figure}}
\floatstyle{plain}\newfloat{myalg}{H}{algs}[section]
\floatname{myalg}{}
\usepackage[letterpaper,
            colorlinks,linkcolor=ForestGreen,citecolor=ForestGreen,
            backref,
            bookmarks,bookmarksopen,bookmarksnumbered]
           {hyperref}

\input{notations}

\begin{document}
\title{Iterative Row Sampling
}

\author{
  Mu Li\\
  CMU\\
  \texttt{muli@cs.cmu.edu}\\
\and
  Gary L. Miller\\
  CMU\\
  \texttt{glmiller@cs.cmu.edu}\\
\and
  Richard Peng\\
  CMU\\
  \texttt{yangp@cs.cmu.edu}\\
}


\maketitle

\input{abstract}


\input{intro}

\input{overview}

\input{prelim}

\input{algo2}

\input{pnorm}




\begin{spacing}{0.7}
  \begin{small}
    \bibliographystyle{alpha}
    \input{Paper.bbl}

  \end{small}
\end{spacing}

\begin{appendix}

\input{leverageproofs}
\input{rowcombineproofs}
\end{appendix}

\end{document}

%% file: notations.tex
\newcommand{\nblock}{n_{b}}
\newcommand{\mata}{\mathbf{A}}
\newcommand{\matashort}{\mathbf{A}\textrm{\tiny$\downarrow$}} 
\newcommand{\vecashort}{\mathbf{a}\textrm{\tiny$\downarrow$}}

\newcommand{\mataapprox}{\tilde{\mathbf{A}}}
\newcommand{\matb}{\mathbf{B}}
\newcommand{\matq}{\mathbf{Q}}

\newcommand{\Ncal}{\mathcal{N}}

\newcommand{\vecb}{\mathbf{b}}

\newcommand{\matc}{\mathbf{C}}
\newcommand{\matd}{\mathbf{D}}

\newcommand{\matu}{\mathbf{U}}
\newcommand{\matuapprox}{\tilde{\mathbf{U}}}

\newcommand{\veczero}{\mathbf{0}}
\newcommand{\matzero}{\mathbf{0}}

\newcommand{\mati}{\mathbf{I}}
\newcommand{\matproj}{\mathbf{P}}

\newcommand{\veca}{\mathbf{a}}

\newcommand{\vecp}{\mathbf{p}}

\newcommand{\vecv}{\mathbf{v}}

\newcommand{\vecx}{\mathbf{x}}
\newcommand{\vecy}{\mathbf{y}}
\newcommand{\vecz}{\mathbf{z}}

\newcommand{\leverage}{\tau}
\newcommand{\leveragev}{\boldsymbol{\tau}}
\newcommand{\approxleverage}{\tilde{{\tau}}}
\newcommand{\approxleveragev}{\tilde{\boldsymbol{\tau}}}

\newcommand{\approxleveragep}{\tilde{\tau}(p)}

\newcommand{\nnz}{\mathrm{nnz}}

\newcommand{\poly}{\mathrm{poly}}
\newcommand{\prob}[1]{\mathrm{Pr}\left( #1 \right)}

\newcommand{\str}[2]{\mathrm{STR}_{#1}(#2)}

\newcommand{\projerror}{R}

\newcommand{\trace}[1]{\textrm{tr}\left[ #1 \right]}

\newcommand{\defequal}{\stackrel{\mathrm{def}}{=}}

\newcommand{\vertiii}[1]{{\left\vert\kern-0.15ex\left\vert\kern-0.15ex\left\vert #1 
\right\vert\kern-0.15ex\right\vert\kern-0.15ex\right\vert}}
\newcommand{\nbr}[1]{\left\|#1\right\|}
\newcommand{\rbr}[1]{\left(#1\right)}
\newcommand{\sbr}[1]{\left[#1\right]}

\newcommand{\invsqr}{{\dag \frac{1}{2}}}

\renewcommand{\Re}{\mathbb{R}}

%% file: abstract.tex
\begin{abstract}
There has been significant interest and progress recently in
algorithms that solve regression problems involving tall and thin
matrices in input sparsity time.
These algorithms find shorter equivalent of a $n \times d$ matrix
where $n \gg d$, which allows one to solve a $\poly(d)$ sized
problem instead.
In practice, the best performances are often obtained by invoking
these routines in an iterative fashion.
We show these iterative methods can be adapted to give
theoretical guarantees comparable and better than the
current state of the art.

Our approaches are based on computing the importances of the rows,
known as leverage scores, in an iterative manner.
We show that alternating between computing a short matrix estimate
and finding more accurate approximate leverage scores leads to a series of
geometrically smaller instances.
This gives an algorithm that runs in
$O(\nnz(A) + d^{\omega + \theta} \epsilon^{-2})$ time for any $\theta > 0$,
where the $d^{\omega + \theta}$ term is comparable to the cost of solving
a regression problem on the small approximation.
Our results are built upon the close connection between
randomized matrix algorithms, iterative methods, and graph sparsification.



\end{abstract}




%% file: intro.tex
\section{Introduction}
\label{sec:intro}

Least squares and $\ell_p$ regression are among the most common
computational linear algebraic operations.
In the simplest form, given a matrix $\mata$ and a vector
$\vecb$, the regression problem aims to find $\vecx$ that
minimizes:
\begin{align*}
\nbr{\mata \vecx - \vecb}_p
\end{align*}
Where $\nbr{\cdot}_p$ denotes the $p$-norm of a vector,
aka.~$\nbr{\vecz}_p = (\sum_i |z_i|^p)^{1/p}$.
The case of $p = 2$ is equivalent to the problem of solving the positive
semi-definite linear system $\mata^T \mata$ \cite{Strang93},
and is one of the most extensively studied algorithmic question.
Over the past two decades, it was shown that $\ell_1$ regression
has good properties in recovering structural information \cite{Candez06Survey}.
These results make regression algorithms a key tool in data analysis,
machine learning, as well as a subroutine in other algorithms.


The ever growing sizes of data raises the natural question
of algorithmic efficiency of regression routines.
In the most general setting, the answer is far from satisfying with
the only general purpose tool being convex optimization.
When $\mata$ is $n \times d$, the state of the theoretical
runtime is about $O((n+d)^{3/2}d)$ \cite{Vaidya89}.
In fact, even in the $\ell_2$ case, the best general purpose
algorithm takes $O(nd^{\omega - 1})$ time where
$\omega \approx 2.3727$ \cite{Vassilevskawilliams12}.
Both of these bounds take more than quadratic time, and more
prohibitively quadratic space, making them unsuitable for modern data
where the number of non-zeros in $\mata$, $\nnz(\mata)$ is often
$10^9$ or more.
As a result, there has been significant interest in either first-order
methods with low per-step cost \cite{Nesterov07,ClarksonHW12},
or faster algorithms taking advantage of additional structures of $\mata$.


One case where significant runtime improvements are possible
is when $\mata$ is tall and thin, aka.~$n \gg d$.
They appear in applications involving many data points in a smaller
number of dimensions, or a few objects on which much data have
been collected.
These instances are sufficiently common that experimental speedups for
finding QR factorizations of such matrices have been studied in the
distributed~\cite{SongLHD10, AgulloCDHL10}
and MapReduce settings~\cite{ConstantineG11}.
Evidences for faster algorithms are perhaps more clear
in the $\ell_2$ setting, where finding $\vecx$ is equivalent to a
linear system solve involving the $d \times d$ matrix $\mata^T \mata$.
When $n \gg d$, the cost of inverting this matrix, $O(d^{\omega})$
is less than the cost of examining the non-zeros in $\mata$.

Faster algorithms for approximating $\mata^T \mata$ were first studied
in the setting of approximation matrix multiplication
~\cite{DrineasKM04a,DrineasKM04b,DrineasKM04c}.
Subsequent approaches were based on finding a shorter matrix
$\matb$ such that solving a regression problem on $\matb$ 
leads to a similar answer \cite{DrineasMM06,DasguptaDHKM09}.
The running time of these routines were also gradually reduced
\cite{Magdonismail10,DrineasMMW11,ClarksonDMMMW12},
leading to algorithms that run in input sparsity time\cite{ClarksonW12,MahoneyM12}.
These algorithms run in time proportional to the number of non-zeros
in $\mata$, $\nnz(\mata)$, plus a $\poly(d)$ term.

An approach common to these algorithms is that they aim
to reduce $\mata$ to $\poly(d)$ sized approximation
using a single transformation.
This transformation is performed in $O(\nnz(\mata))$ time, after which
the problem size only depends on $d$, giving the $\poly(d)$ term.
This is done by either obtaining high quality sampling probabilities
\cite{DrineasMMW11,ClarksonDMMMW12}, or by directly creating $\matb$
via. a randomized transform \cite{ClarksonW12,MahoneyM12,NelsonN12}.
These algorithms are appealing due to simplicity, speed, and that they can be
adapted naturally in the streaming setting.
On the other hand, experimental works have shown that practical
performances are often optimized by applying higher error variants
of these algorithms in an iterative fashion \cite{AvronMT10}.

In this paper, we design algorithms motivated by these practical adaptations
whose performances match or improve over the current best.
Our algorithms  construct $\matb$ containing $\poly(d)$ rows of $\mata$
and run in $O(\nnz(\mata) + d^{\omega + \theta})$ time.
Here the last term is due to computing inverses and change of basis matrices,
and is a lower order term since regression routines involving $d \times d$
matrices take at least $d^{\omega}$ time.
In Table~\ref{table:tableall} we give a quick comparison of our results with
previous ones in the $\ell_2$ and $\ell_1$ settings can be found .
These two norms encompass most of the regression problems
solved in practice \cite{Candez06Survey}.
To simplify the comparison, we do not distinguish between
$\log{d}$ and $\log{n}$, and assume that $\mata$ has full column rank.
We will also omit the big-O notation along with factors of
$\epsilon$ and $\theta$.

\renewcommand{\arraystretch}{1.2}
\begin{table}[ht]
\begin{center}
\begin{tabular}{|l|c|c|c|c|}
\hline
&\multicolumn{2}{|c|}{\textbf{$\ell_2$}}
&\multicolumn{2}{|c|}{\textbf{$\ell_1$}}\\
\cline{2-5}
& Runtime & \# Rows
& Runtime & \# Rows\\
\hline
Dasgupta et al. \cite{DasguptaDHKM09}
& \multicolumn{2}{|c|}{-}
& $nd^5\log{d}$ & $d^{2.5}$\\
\hline
Magdon-Ismail \cite{Magdonismail10}
& $nd^{2} / \log{d}$ & $d\log^{2}d$
& \multicolumn{2}{|c|}{-} \\
\hline
Sohler \& Woodruff \cite{SohlerW11}
& \multicolumn{2}{|c|}{-}
& $n d^{\omega  - 1 + \theta}$ & $d^{3.5}$\\
\hline 
Drineals et al. \cite{DrineasMMW11}
& $nd\log{d}$ & $d\log{d}$
& \multicolumn{2}{|c|}{-} \\
\hline
Clarkson et al. \cite{ClarksonDMMMW12}
& \multicolumn{2}{|c|}{-}
& $nd\log{d}$ & $d^{4.5} \log^{1.5}{d}$\\
\hline
Clarkson \& Woodruff \cite{ClarksonW12}
& $\nnz(\mata)$ & $d^2\log{d}$
& $\nnz(\mata) + d^{7} $ & $d^{8} \poly(\log{d})$\\
\hline
Mahoney \& Meng \cite{MahoneyM12}
& $\nnz(\mata)$ & $d^2$
& $\nnz(\mata) \log{n} + d^{8}$ & $d^{3.5}$\\ 
\hline
Nelson \& Nguyen \cite{NelsonN12} & $\nnz(\mata)$ & $d^{1 + \theta}$
& \multicolumn{2}{|c|}{Similar to \cite{ClarksonW12}~and~\cite{MahoneyM12}}\\
\hline
\textbf{This paper}
& $\nnz(\mata) + d^{\omega + \theta}$ & $d \log{d}$
& $\nnz(\mata) + d^{\omega + \theta}$ & $d^{3.66}$\\
\hline
\end{tabular}
\caption{Comparison of runtime and size of $\matb$ for
$\ell_2$ and $\ell_1$, $\theta$ is any constant that's $> 0$}
\label{table:tableall}
\end{center}
\end{table}

As with previous results, our approaches and bounds for
$\ell_2$ and $\ell_p$ are fairly different.
We will state them in more details and give a more
detailed comparison with previous results in Section~\ref{sec:overview}.
The key idea that drives our algorithms is that a constant factor
reduction of problem size suffices for a linear time algorithm.
This is a much weaker requirement than reducing directly
to $\poly(d)$ sized instances, and allows us to reexamine
statistical projections with weaker guarantees.
In the $\ell_2$ setting, projections that do not
even preserve the column space of $\mata$ can still give good
enough sampling probabilities.
For the $\ell_p$ setting, estimating the probabilities
in the `wrong' norm (e.g.~$\ell_2$) still leads to significant reductions.
Most of the subroutines that we'll use have either
been used as the final error correction step
\cite{DasguptaDHKM09,ClarksonDMMMW12,ClarksonW12},
or are known in folklore.
However, by combining these tools with techniques originally developed
in graph sparsification and combinatorial preconditioning \cite{KoutisLP12},
we are able to convert them into much more powerful algorithms.
A consequence of the simplicity of the routines used is
that we obtain a smaller number of rows in $\matb$ in the $\ell_2$
setting, as well as a smaller running time in the $\ell_p$ setting.
We believe these reductions in the $\poly(d)$ term are crucial for
closing the gap between theory and practice of these algorithms.

%% file: overview.tex
\section{Overview}
\label{sec:overview}

We start by formalizing the requirements needed for $\matb$ to be a
good approximation to $\mata$.
In the $\ell_2$ setting it is similar to $\matb^T \matb$ being an approximation
to $\mata^T\mata$, but looking for $\matb$ instead of $\matb^T \matb$ has
the advantage of being extendible to $\ell_p$ norms \cite{DasguptaDHKM09}.
The requirement for $\matb$ is:
\begin{align*}
(1 - \epsilon) \|\mata \vecx\|_p
\leq \|\matb \vecx\|_p
\leq (1 + \epsilon) \|\mata \vecx\|_p,
\qquad \forall \vecx \in \Re^{d}
\end{align*}
Finding such a $\matb$ is equivalent to reducing the size of a
regression problem involving $\mata$ since:
\begin{align*}
\min_{\vecx} \|\mata \vecx - \vecb\|_p
= \min_{\vecx} \nbr{[\mata, \vecb]
  \begin{bmatrix}
    \vecx \\ -1
  \end{bmatrix}}_p
\end{align*}
This means finding a shorter $(1 \pm \epsilon)$ approximation to the
$n \times (d + 1)$ matrix $[\mata, \vecb]$, and solving a regression
problem on this approximation gives a solution within
$1 + O(\epsilon)$ of the minimum.

Row sampling is one of the first studied approaches for finding
such $\matb$ \cite{DrineasMM06,Magdonismail10,DasguptaDHKM09}.
It aims to build $\matb$ consisting of a set of rescaled
rows of $\mata$ chosen according to some distribution.
While it appears to be a even more restrictive way of generating $\matb$,
it nevertheless leads to a row count within a factor of $\log{d}$ of the
best known bounds~\cite{BatsonSS09,BoutsidisDM11}.
In $\ell_2$, there exists a distribution that produces with high probability
a good approximation $\matb$ with $O(d \log{d})$ rows
~\cite{AhlswedeW02,RudelsonV07,VershyninNotes,Harvey11notes};
while under $\ell_p$ norm, $\poly(d)$ rows is also known~\cite{DasguptaDHKM09}.
It was first shown that row sampling can speed up $\ell_2$
this can be viewed as a small subset that preserves most of the structure.
These smaller equivalents have been studied as
coresets under a variety of objectives~\cite{BadoiuHP02,AgarwalHV05}.
However, various properties of the $\ell_p$ norm, especially
in the case of $p = 2$, makes row sampling a more specialized instance.

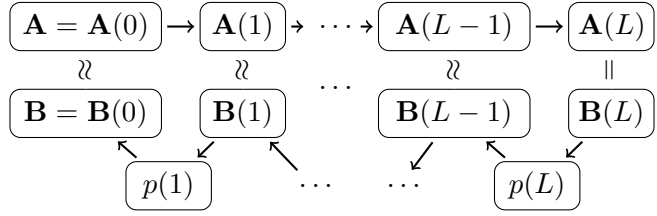
\begin{figure}[t!]
\begin{center}

\begin{tikzpicture}[x=1cm, y=1cm]
\usetikzlibrary{arrows,positioning} 

\tikzset{
    roundrect/.style={
           rectangle,
           rounded corners,
           draw=black,
           text width=25,
           minimum height = 0.6,
           text centered},
    roundrectwide/.style={
           rectangle,
           rounded corners,
           draw=black,
           text width=50,
           minimum height = 0.6,
           text centered},
    approxstretch/.style={
           draw=black,
           text width=10,
           minimum height = 0.6,
           text centered},
    blank/.style={
           text width=10},
    arrow/.style={
           ->,
           thick,
           shorten <=2pt,
           shorten >=2pt,},
    empty/.style={white}
}

\draw node[roundrectwide](aa){$\mata = \mata(0)$};
\draw node[roundrect, right=0.5 of aa](ab){$\mata(1)$}; 
\draw node[blank, right = 0.3 of ab](amid) {$\ldots$}; 
\draw node[roundrectwide, right = 0.3 of amid](ay){$\mata(L - 1)$}; 
\draw node[roundrect, right=0.5 of ay](az) {$\mata(L)$}; 
\draw (aa) edge[arrow] (ab);
\draw (ab) edge[arrow] (amid);
\draw (amid) edge[arrow] (ay);
\draw (ay) edge[arrow] (az);
\draw node[roundrectwide, below=0.5 of aa](ba) {$\matb = \matb(0)$};
\draw node[roundrect, below=0.5 of ab](bb) {$\matb(1)$};
\draw node[blank, below = 0.5 of amid](bmid){$\ldots$}; 
\draw node[roundrectwide, below = 0.5 of ay](by) {$\matb(L - 1)$}; 
\draw node[roundrect, below=0.5 of az](bz) {$\matb(L)$}; 
\draw node[blank, below=0.5 of bb](dummyb) {}; 
\draw node[blank, below=0.5 of by](dummyy) {}; 
\draw node[blank, below=0.5 of bz](dummyz) {}; 
\draw node[roundrect, left=0.1 of dummyz](lz) {$p(L)$}; 
\draw node[blank, right= 0.3 of dummyb](lmidl){$\ldots$}; 
\draw node[blank, left=0.1 of dummyy](lmidr) {$\ldots$}; 
\draw node[roundrect, left=0.1 of dummyb](lb) {$p(1)$}; 
\draw (aa) edge[empty] node[rotate=90,black] {$\approx$}(ba);
\draw (ab) edge[empty] node[rotate=90,black] {$\approx$}(bb);
\draw (ay) edge[empty] node[rotate=90,black] {$\approx$}(by);
\draw (az) edge[empty] node[rotate=90,black] {$=$}(bz);
\draw (bz) edge[arrow] (lz);
\draw (by) edge[arrow] (lmidr);
\draw (bb) edge[arrow](lb);
\draw (lz) edge[arrow] (by);
\draw (lmidl) edge[arrow] (bb);
\draw (lb) edge[arrow] (ba);
\end{tikzpicture}
\end{center}
\caption{Main workflow of our algorithms when viewed as an
iterative process.
Sequence of gradually smaller matrices generated are on top,
and the computed sampling probabilities and resulting
approximations are below.}
\label{fig:flowchart}
\end{figure}

The main framework of our algorithm is iterative in nature and relies on
the two-way connection between row sampling and estimation
of sampling probabilities.
A crude approximation to $\mata$, $\mata'$ allows us to compute equally
crude approximations of sampling probabilities, while such probabilities
in turn lead to higher quality approximations.
The computation of these sampling probabilities can in turn be sped up
using a high quality approximation of $\mata'$.
Our algorithm is based on observing that as long as $\mata'$ has smaller
size, we have made enough progress for an iterative algorithm.
A single step in this algorithm consists of computing a small but crude
approximation $\mata'$, finding a higher quality approximation to $\mata'$,
and using this approximation to find estimates of sampling probabilities
of the rows of $\mata$.
This leads to a tail-recursive process that can also be viewed as an iterative
one where the calls generate a sequence of gradually shrinking matrices,
and sampling probabilities are propagated back up the sequence.
An example of such a sequence is given in Figure \ref{fig:flowchart}.

We will term the creation of the coarse approximation as reduction,
and the computation of the more accurate approximation based on it
recovery.
As in the figure, we will label the matrices $\mata$ that we generate,
as well as their approximations using the indices $(l)$.

\begin{itemize}
        \item reduction:        creates a smaller version of $\mata(l)$,
$\mata(l + 1)$ with fewer rows either by a projection or a coarser row
sampling process. Equilvaent to moving rightwards in the diagram.
        \item recovery: finds a small, high quality approximation
of $\mata(l)$, $\matb(l)$ using information obtained from $\mata(l)$, $\mata(l + 1)$,
and $\matb(l + 1)$.
This is done by estimating leverage scores $\vecp(l)$ and is equivalent
to moving leftwards in the diagram.
\end{itemize}

Both our $\ell_2$ and $\ell_p$ algorithms can be
viewed as giving reduction and recovery routines.
In the $\ell_2$ setting our reduction step consists of a simple
random projection, which incurs a fairly large distortion and may not
even preserve the null space.
Our key technical components in Section~\ref{sec:algo2} show
that one-sided bounds on these projections are sufficient for recovery.
This allows us to set the difference incurred by the reduction to
$\kappa = d^{\theta}$ for a arbitrarily small $\theta > 0$,
while obtaining a reduction factor of $\kappa^{O(1)} = d^{O(\theta)}$.
This error is absorbed by the sampling process, and does not
accumulate across the iterations.

\begin{theorem}
\label{thm:rowsamplel2}
Given a $n \times d$ matrix $\mata$ along with failure probability
$\delta = d^{-c}$ and allowed error $\epsilon$.
For any constant $\theta > 0$,
we can find in $O( \nnz(A) + d^{\omega + \theta} \epsilon^{-2})$ time,
with probability at least $1 - \delta$, a matrix $\matb$ consisting of
$O(d\log{d}\epsilon^{-2})$ rescaled rows of $\mata$ such that
$$(1-\epsilon)\|\mata \vecx\|_2 \le \|\matb \vecx\|_2 \le (1+\epsilon) \|\mata \vecx\|_2$$
for all vectors $\vecx \in \Re^{d}$.
\end{theorem}

This bound improves the $O(d^2)$ rows obtained in the first results
with input-sparsity runtime \cite{ClarksonW12}, and matches
the best bound known using oblivious projections \cite{NelsonN12},
which was obtained concurrently.
A closer comparison with \cite{NelsonN12} shows that
our bounds does not have a factor of $\epsilon^{-1}$ on the leading term
$\nnz(\mata)$, but has worse dependencies on $\theta$.

For $\ell_p$ norms, we show that significant size
reductions can be made if we perform row sampling using sampling
probabilities obtained in a different norm.
Specifically, if $\mata(i)$ has $n(i)$ rows, $\mata(i + 1)$ has
$O(n(i)^{c_p} \poly(d))$ where $c_p < 1$ if the intermediate
norm $\ell_{p'}$ is chosen appropriately.
This means the number of rows will reduce doubly exponentially
as we iterative, and quickly becomes $O(\poly(d))$.

This allows us to invoke our algorithms from Section~\ref{sec:algo2} ,
as well as $\ell_p'$ approximations under different norms to compute
these probabilities.
The analysis is also more direct as such samples have stronger
guarantees than randomized projections,
We can set $\kappa$ to a constant, and recover an
approximation to $\mata$ after each iteration instead of going
gradually back up the sequence of matrices.

Our projection and recovery methods are similar to the ones
used to for increasing the accuracy of $\ell_1$ row sampling
in \cite{ClarksonDMMMW12}.
However, to our knowledge, our result is the first that uses
$\ell_2$ row sampling as the primary routine.
This leads to the first algorithms for $p \neq 1, 2$ that do
not use ellipsoidal rounding.
In Section~\ref{subsec:oneshot} we present a one step variant
that computes sampling probabilities under the $\ell_2$ norm.
It gives $\matb$ with about $d^{\frac{4}{p}}$ rows when $p \leq 2$
and $d^{\frac{3p - 2}{4 - p}}$ rows when $2 \leq d < 4$.
We can further iterate upon this algorithm, and compute sampling
probabilities under $\ell_{p'}$ norm for some $p'$ between $2$ and $p$.
A two-level version of this algorithm for $\ell_1$ is analyzed
in Section~\ref{subsec:again}, giving the following:
\begin{theorem}
\label{thm:rowsamplel1}
Given a $n \times d$ matrix $\mata$ along with failure probability
$d^{-c}$ and allowed error $\epsilon$.
For any constant $\theta > 0$,
we can find in $O( \nnz(A) + d^{\omega + \theta} \epsilon^{-2})$ time,
with probability at least $1 - d^{-c}$, a matrix $\matb$ consisting of
$O(d^{4\sqrt{2} - 2 + \theta})$ rescaled rows of $\mata$ such that
$$(1-\epsilon)\nbr{\mata \vecx}_1  \le \nbr{\matb \vecx}_1 
        \leq (1+ \epsilon) \nbr{\mata \vecx}_1$$
for all vectors $\vecx \in \Re^d$.
\end{theorem}
This method readily leads to $\matb$ with $\poly(d)$ rows when $p \ge 4$,
and fewer rows than the above bound when $1 \le p < 4$.
However, such extensions are limited by the discontinuity between
bounds on the sampling process in the $\ell_2$
~\cite{AhlswedeW02,RudelsonV07,VershyninNotes,Harvey11notes}
and $\ell_p$ settings~\cite{DasguptaDHKM09}.
As a result, we only show the algorithm for $\ell_1$
in order to simplify the presentation.

An additional strength of our approach is that the randomized routines
used hold with high probability.
Most of the earlier results that run in time nearly-linear in the size of
$\mata$ have a constant success probability instead, and will require
boosting to improve this probability.
Also, as our algorithm is row sampling based, each row in our output
is a scaled copy of some row of the original matrix.
This means specialized structure for rows of $\mata$ are likely to be
preserved in the smaller regression problem instance.
Our results also show a much tighter connection between $\ell_2$
and $\ell_p$ row sampling, namely that finding good $\ell_2$
approximations alone is sufficient for iterative reductions in matrix size.

The main drawback of our algorithm in the $\ell_2$ setting
is that it does not immediately extend to computing low-rank approximations.
The method given in \cite{ClarksonW12} relies crucially on
first transform being oblivious, although our algorithm can
be incorporated in a limited way as the second step.
Also, our algorithms for $\ell_p$ row-sampling in Section~\ref{sec:pnorm}
invokes concentration bounds from~\cite{DasguptaDHKM09} in a
black-box manner, even though our sampling probabilities obtained
by scaling up probabilities related to $\ell_2$.
We believe investigating the possibilities of extending our approaches to
low-rank approximations and obtaining tighter concentration are natural
directions for future work.

%% file: prelim.tex
\section{Preliminaries}
\label{sec:prelim}

We begin by stating key notations and definitions that we will use
for the rest of this paper.
We will use $\|\vecx\|_p$ to denote the $\ell_p$ norm of a vector.
The two values of $p$ that we'll use are $p=1$ and $p=2$,
which correspond to $\|\vecx\|_1 = \sum_{i} |x_i|$
and $\|\vecx\|_2 = \sqrt{\sum_{i} x_i^2}$.
For two vectors $\vecx$ and $\vecy$, $\vecx \geq \vecy$ means
$\vecx$ is entry-wise greater or equal to $\vecy$, aka.~$\vecx_i \geq \vecy_i$ for all $i$.

For a matrix $\mata$, we use $\mata_{i*}$, or $\veca_i$
to denote the $i$\textsuperscript{th} row of $\mata$, and
$\mata_{*j}$ to denote its $j$\textsuperscript{th} column.
Note that if $\mata \in \Re^{n \times d}$, $\veca_i$ is a row vector
of length $d$.
We will also use the generalized $p$-norm $\vertiii{\cdot}_p$
of a matrix, which essentially treats all entries of the matrix as
a single vector.
Specifically, $\vertiii{\mata}_p = (\sum_{ij} |\mata_{ij}|^{p})^{1/p}$.
When $p = 2$, it is known as the Frobenius norm, $\|\cdot\|_F$.

A matrix $\matc$ is positive semi-definite
if all its eigenvalues are non-negative,
or equivalently $\vecx^T \matc \vecx \geq 0$
for all vectors $\vecx$.
Since $\vecx^T (\mata^T \mata) \vecx = \|\mata \vecx\|_2^2$,
$\mata^T \mata$ is positive semi-definite for any $\mata$.
Similarity between matrices is defined via. a partial order on matrices.
Given two matrices $\matc_1$ and $\matc_2$, $\matc_1 \preceq \matc_2$
denotes that $\matc_2 - \matc_1$ is positive semi-definite.
The connection between this notation and row sampling is clear
in the case of $\ell_2$, specifically $\mata^T \mata \preceq \matb^T \matb$
is equivalent to $\|\mata \vecx\|_2 \leq \|\matb \vecx\|_2$.

We will also define the pseudoinverse of $\matc$, $\matc^{\dag}$
as the linear operator that's zero on the null space of $\matc$,
while acting as its inverse on the rank space.
For operators that act on the same space, spectral orderings of
pseudoinverses behaves the same as with scalars.
Specifically, if $\matc_1$ and  $\matc_2$ have the same
null space and $\matc_1 \preceq \matc_2$,
then $\matc_2^{\dag} \preceq \matc_1^{\dag}$.
Given a subspace of $\Re^{d}$, an orthogonal projector onto it,
$\matproj$ is a symmetric positive-semidefinite matrix taking
vectors into their projection in this space.
For example, if this space is rank space of some positive semi-definite
matrix $\matc$, then an orthogonal projection operator is given
by $\matc \matc^{\dag}$. 

Our algorithms are designed around the following algorithmic fact:
for any norm $p$ and any matrix $\mata \in \Re^{n \times d}$,
there exist a distribution on its rows such that sampling $\poly(d)$
entries from this distribution and rescaling them gives $\matb$
such that with probability at least $1 - d^{-c}$:
\begin{align*}
(1 - \epsilon) \|\mata \vecx\|_p
\leq \|\matb  \vecx\|_p
\leq (1 + \epsilon) \|\mata \vecx\|_p,
\qquad \forall \vecx \in \Re^{d}.
\end{align*}
This sampling process can be formalized in several ways, leading to
similar results both theoretically and experimentally \cite{IpsenW12}.
We will treat it as a blackbox $\textsc{Sample}(\mata, \vecp)$
that takes a set of probabilities over the rows of $\mata$ and
samples them accordingly.
It keeps row $i$ or $\mata$ with probability $\min \{ 1, p_i\}$, and
rescales it appropriately so the expected value of this row is preserved.
The two key properties of $\textsc{Sample}(\mata, \vecp)$ that we
will use repeatedly are:
\begin{itemize}
\item It returns $\matb$ with at most $O(|\vecp|_1)$ rows.
\item Its running time can be bounded by $O(n + |\vecp|_1 \log{n})$.
\end{itemize}


The convergence of sampling relies on matrix Chernoff bounds,
which can be viewed as generalizations of single variate
concentration bounds.
Necessary conditions on the probabilities can be formalized in several
ways, with the most common being statistical leverage scores.
Although these values have been studied in statistics,
their use in algorithms is more recent.
To our knowledge, their first use in a limited row sampling setting
was in spectral sparsification of graphs \cite{SpielmanS08}.
The most general definition of $p$-norm leverage scores is based on
the row norms of a basis of the column space of $\mata$.
However, significant simpliciations are possible when $p = 2$,
and this alternate view is crucial for in our algorithm.
As a result, we will state the relevant convergence results
for $\textsc{Sample}$ separately in
Sections~\ref{sec:algo2}~and~\ref{sec:pnorm}.

They show that statistical leverage scores are closely
associated the probabilities needed for row sampling,
and give algorithms that efficiently approximate these values.
We will also formalize an observation implicit in previous results
that both the sampling and estimation algorithms are very robust.
The high error-tolerance of these algorithms makes them ideal
as core routines to build iterative algorithms upon.

One issue with the various concentration bounds that we will prove
is that they hold with high probability in $d$.
That is, the fail with probability $1 - d^{-c}$ for some constant $c$.
In cases where $n \gg \poly(d)$, this will prevent
us from taking a union bound over many sampling steps.
However, it can be shown that in such cases, padding all sampling
probabilities with $1 / \poly(d)$ in the sampling process will narrow
the key steps back down to $\poly(d)$ ones.
This leads to a matrix with $O(\nnz(\mata) / \poly(d))$ rows,
which can in turn be handled in $O(\nnz(\mata))$ time using
routines that run in $O(n \poly(d))$ time (e.g. \cite{DasguptaDHKM09}).
Therefore for the rest of this paper we will assume $n = \poly(d)$.


%% file: algo2.tex
\section{Iterative Row Sampling for $\ell_2$}
\label{sec:algo2}


We start by presenting our algorithm for computing row sampling
in the $\ell_2$ setting.
Crucial to our approach is the following basis-free definition of
statistical leverage scores of $\leveragev$: 
\begin{align*}
\leverage_i
\defequal \veca_i (\mata^T \mata)^{+} \veca_i^{T}, \quad \textrm{for } i=1,\ldots,n,
\end{align*}
where $\veca_i$ is the $i$-th row of $\mata$.

To our knowledge, the first near tight bounds for row sampling using
statistical leverage scores were given in \cite{AhlswedeW02},
and various extensions and simplifications were made since
\cite{RudelsonV07,VershyninNotes, Harvey11notes,AvronT11}.
They can be stated as follows:
\begin{lemma}
\label{lem:samplel2}
If $\approxleveragev$ is a set of probabilities such that
$\approxleveragev \geq \leveragev$, then for any constants
$c$ and $\epsilon$, there exists a function 
$\textsc{Sample}(\mata, O(\log{d}, \epsilon) \approxleveragev)$ which returns $\matb$
containing 
$O(\log d \nbr{\approxleveragev}_1\epsilon^{-2})$ rows and satisfying 
\begin{align*}
(1 - \epsilon) \| \mata \vecx \|_2
\leq \| \matb \vecx \|_2
\leq (1 + \epsilon) \| \mata \vecx \|_2,
\qquad \forall \vecx\in\Re^d
\end{align*}
with probability at least $1 - d^{-c}$.
\end{lemma}

The importance of statistical leverage scores can be reflected
in the following fact, which implies that we can obtain $\matb$
with $O(d\log{d})$ rows.
\begin{fact}
\label{fact:leveragesum} (see e.g.~\cite{SpielmanS08}) Given $n\times d$ matrix $\mata$, and let $\leveragev$ be the
leverage score w.r.t.~$\mata$. Assume $\mata$ has rank $r$, then 
\begin{align*}
\sum_{i = 1}^n \leverage_i = r \leq d
\end{align*}
\end{fact}

Although it is tempting to directly obtain high quality
approximations of the leverage scores, their computation
also requires a high quality approximation of $\mata^T \mata$,
leading us back to the original problem of row sampling.
Our way around this issue relies on the robustness of concentration
bounds such as Lemma~\ref{lem:samplel2}.
Sampling using even crude estimates on leverage scores can lead
to high quality approximations
\cite{DasguptaDHKM09,DrineasM10, DrineasMMS11,DrineasMMW11,AvronT11}.
Therefore, we will not approximate $\mata^T \mata$ directly,
and instead obtain a sequence of gradually better approximations.
The need to compute sampling probabilities using crude approximations
leads us to define a generalization of statistical leverage scores.


\subsection{Generalized Stretch and its Estimation}
\label{subsec:generalizedstretch}

The use of different matrices to upper bound stretch has
found many uses in combinatorial preconditioning, where it's
termed stretch \cite{SpielmanW09,KoutisLP12,DrineasM10}.
We will draw from them and term our generalization
of leverage scores \textbf{generalized stretch}.
We will use $\str{\matb}{\veca_i}$ to denote the approximate
leverage score of row $i$ computed as follows:
\begin{align}
\str{\matb}{\veca_i}
\defequal \veca_i (\matb^T \matb) ^{\dag} \veca_i
\end{align}
Under this definition, the original definition of statistical leverage
score $\leverage_i$ equals to $\str{\mata}{\veca_i}$.
We will refer to $\matb$ as the reference used to compute stretch.
It can be shown that when $\matb_1$ and $\matb_2$ are reasonably
close to each other, stretch can be used as upper bounds for
leverage scores in a way that satisfies Lemma~\ref{lem:samplel2}.
\begin{lemma}
\label{lem:referenceswitch}
If $\matb_1$ and $\matb_2$ satisfies:
\begin{align*}
\frac{1}{\kappa} \matb_1^T \matb_1
\preceq  \matb_2^T \matb_2
\preceq \matb_1^T \matb_1
\end{align*}
Then for any vector $\vecx$ we have:
\begin{align*}
\str{\matb_1}{\vecx}  \leq \str{\matb_2}{\vecx} \leq \kappa \str{\matb_1}{\vecx}
\end{align*}
\end{lemma}
The proof will be shown in Appendix~\ref{sec:leverageproofs}.

The stretch notation can also be extended to a set of rows, aka.~a matrix.
If $\mata$ is a matrix with $n$ rows, $\str{\matb}{\mata}$ denotes:
\begin{align}
\str{\matb}{\mata}
\defequal \sum_{i = 1}^{n} \str{\matb}{\veca_{i}}.
\end{align}
This view is useful as it allows us to write stretch
as the $\ell_2$ norm of a vector, or more generally
the stretch of a set of rows as the Frobenius norm of a matrix.
\begin{fact}
\label{fact:stretchclosedform}
The generalized stretch of the $i^{th}$ row of $\mata$ w.r.t $\matb$ equals
to its $\ell_2^2$ norm under the transformation $(\matb \matb^T)^\invsqr$:
\begin{align*}
\str{\matb}{\veca_i}
= \| (\matb \matb^T)^\invsqr \veca_i^T \|_2^2
= \| \veca_i (\matb \matb^T)^\invsqr \|_2^2
\end{align*}
and the total stretch of all rows is
\begin{align*}
\str{\matb}{\mata}
=  \| (\matb \matb^T)^\invsqr \mata^T \|_F^2
= \|\mata (\matb \matb^T)^\invsqr \|_F^2
\end{align*}
\end{fact}

This representation leads to faster algorithms for estimating stretch
using the Johnson-Lindenstrauss transform.
This tool is used in a variety of settings from estimating effective
resistances \cite{SpielmanS08} to more generally leverage scores
\cite{DrineasMMW11}.
We will use the following randomized projection theorem:


\begin{lemma}
  \label{lem:jl} (Lemma 2.2 from \cite{dasgupta2003elementary})
  Let $\vecy$ be a unit vector in $\Re^d$. Then for any positive integer $k\le d$, let
  $\matu$ be a $k\times d$ matrix with entries chosen independently from the Gaussian
  distribution $\Ncal(0,1)$. Let $\vecx = \matu \vecy$ and
  $L=\|\vecx\|_2^2$. Then for any $\projerror >1$,
  \begin{enumerate}
  \item \label{part:expectation} $\mathbb E(L) = k$
  \item \label{part:upper} $\prob{L \ge {\projerror k}} <
    \exp\rbr{\frac{k}{2}(1-\projerror+\ln{\projerror})}$
\item \label{part:lower} $\prob{L \le \frac{k}{\projerror }} <
    \exp\rbr{\frac{k}{2}(1-\projerror^{-1}-\ln{\projerror})}$
   \end{enumerate}
  
\end{lemma}
We will also use this lemma in our reduction step to bound the
distortion when rows are combined.
Note that the requirements of Lemma~\ref{lem:samplel2}
and the guarantees of \textsc{Sample} allows our estimates
to have larger error.
This means we can use fewer vectors in the projection, and
scale up the results to correct potential underestimates.
Therefore, we can trade the coefficient on the leading term
$\nnz(\mata)$ with a higher number of sampled row count.
The bound below accounts for both error incurred by $\matb$,
and the larger error caused by this error.

\begin{lemma}
\label{lem:approxstretch}
For any constant $c$, there is a routine
$\textsc{ApproxStr}(\mata, \matb, \kappa, \projerror)$, shown in Algorithm~\ref{alg:approxstretch},
that when given a $n \times d$ matrix $\mata$ where $n = \poly(d)$,
and an approximation $\matb$ with $m$ rows such that:
\begin{align*}
\frac{1}{\kappa} \mata^T \mata \preceq \matb^T \matb \preceq \mata^T \mata
\end{align*}
return in 
$O( (\nnz(A) + d^{2} ) \log_{\projerror}{d}  + (m+d)d^{\omega-1})$
time and upper bounds $\approxleverage_i$ such that with probability at least $1 - d^{-c}$
\begin{enumerate}
\item \label{part:l2stretchupper} for all $i$, $\approxleverage_i \geq \leverage_i$.
\item \label{part:l2stretchsum} $\| \approxleveragev \|_1 \leq O(\projerror^2 \kappa d
  )$. 
\end{enumerate}
\end{lemma}

\subsection{Reductions and Recovery}
\label{sec:rowproj}
Our reduction and recovery processes are based on projecting $\mata$
to one with fewer rows, and moving the estimates on the projection
back to the original matrix.
Our key operation is to combine every $R$ rows into $k$ rows, where
$R$ and $k$ are set to $d^{\theta}$ and $O(c / \theta)$ respectively.
By padding $\mata$ with additional rows of zeros, we may assume
that the number of rows is divisible by $R$.
We will use $\nblock = n/R$ to denote the number of blocks,
and use the notation $\cdot_{(b)}$ to index into the $b$\textsuperscript{th} block.
Our key step is then a $(R, k)$-reduction of the rows:
\begin{definition}
\label{def:reduction}
A $(R, k)$-reduction of $\mata$ describes the following procedure:
\begin{enumerate}
        \item For each block $\mata_{(b)}$, pick $\matu_{(b)}$ to
                be a $k \times R$ random Gaussian matrix
                with entries picked independently from $\mathcal{N}(0, 1)$ and
                compute $\matashort_{(b)} = \matu_{(b)} \mata_{(b)}$.
        \item Concatenate the blocks $\matashort_{(b)}$ together vertically to form $\matashort$.
\end{enumerate}
\end{definition}


We first show that projections preserve the stretch of blocks
w.r.t. $\mata$.
This can be done by bounding the effect of $\matu_{(b)}$ on
the norm of each column of $\mata_{(b)} (\mata^T \mata)^\invsqr$.
It follows directly from properties of the Johnson-Lindenstrauss
projections described in Lemma \ref{lem:jl}, and we'll give
its proof in Appendix~\ref{sec:rowcombineproofs}.

\begin{lemma}
\label{lem:goodapprox}
Assume $R = d^{\theta} \ge e^2$ for some constant $\theta$ and
let $\matashort$ be a $(R, k)$-projection of $\mata$.
For any constant $c > 0$ there exists a constant
$k = O(c / \theta)$ such that 
\begin{align*}
\str{\mata}{\matashort_{(b)}}
 \geq  \frac{k}{R} \str{\mata}{\mata_{(b)}}
\end{align*}
holds for all block $b=1,\ldots,n_b$
with probability at least $1 - d^{-c}$.
\end{lemma}

We next show that we can change the reference from $\mata$
to $\matashort$, and use $\str{\matashort}{\matashort_{(b)}}$
as upper bounds for $\str{\mata}{\matashort_{(b)}}$.
As a first step, we need to relate $\mata^T \mata$ to
$\matashort^T\matashort$.
Since each $\matashort_{(b)}$ is formed by merging rows
of $\matashort_{(b)} = \matu_{(b)} \mata_{(b)}$,
$\matashort_{(b)}^T\matashort_{(b)}$ can
be upper bounded by $\mata_{(b)}^T\mata_{(b)}$ times
a suitable term depending on $\matu_{(b)}$.
We prove the following in Appendix~\ref{sec:rowcombineproofs}.

\begin{lemma}
\label{lem:shortupper}
The following holds for each block $b$:
\begin{align*}
        \matashort_{(b)}^T \matashort_{(b)}
        \preceq  \|\matu_{(b)} \|_F^2 \cdot \mata_{(b)}^T \mata_{(b)}
\end{align*}
\end{lemma}

However, generalized stretches w.r.t.~$\mata$ and $\matashort$
are evaluated under the norms given by the inverses of these
matrices, $(\mata^T\mata)^{+}$ and $(\matashort^T\matashort)^{+}$.
As a result, we need to bound the operator bound between these two
pseudoinverses, which we obtain using the following lemma.

\begin{lemma}
  \label{lem:pseudoinversereverse}

  Let $\matc$ and $\matd$ be symmetric positive semi-definite matrices and let
  $\matproj$ be the orthogonal projection operator onto the range space of
  $\matc$.  Then:
 \begin{align*} \matproj \matc^{+} \matproj \succeq \matproj (\matc + \matd)^{+}
\matproj
 \end{align*}
\end{lemma}

This is straightforward when both $\matc$ and $\matd$ are
full rank, or share the same null space.
However, as pseudo-inverses do not act on the null space,
it is crucial that we're only considering vectors of the form $\veca'_i$.
This Lemma is proven in Appendix \ref{sec:rowcombineproofs}.
Combining it with bounds in the other direction allows us to bound
the distortion caused by switching reference from $\mata$ to $\matashort$.

\begin{lemma}
\label{lem:leverageupper}
For any constant $c$, there exists 
a constant $c'$,
such that with probability at most $1 - d^{-c}$,
we have for each row $i$ of $\matashort$, denoted by $\vecashort_i$, satisfies 
\begin{align*}
c' k R \log d \cdot  \str{\matashort}{\vecashort_i}
\geq \str{\mata}{\vecashort_i}
\end{align*}
\end{lemma}

\Proof Denote by $\mata_{(b)}$ the $b$-th block of $\mata$ and $\matashort_{(b)}$ the
corresponding block in $\matashort$, by Lemma~\ref{lem:shortupper},
\begin{align*}
  \matashort_{(b)}^T \matashort_{(b)} 
  \preceq \nbr{\matu_{(b)}}_F^2\mata_{(b)}^T 
  \mata_{(b)}
  \end{align*}
Since each
$\matu_{(b)}$ consists of $k \times R$ independent random variables chosen from
$\mathcal{N}(0, 1)$, $\| \matu_{(b)} \|_F^2$ is distributed as $\mathcal{N}(0, kR)$.  This
gives:
\begin{align}
\prob{\| \matu_{(b)} \|_F^2 > \ell kR} 
\leq \exp(-\ell)
\end{align}
As $n=\poly(d)$, this probability can
be bounded by $d^{-c}n^{-1}$ for an appropriate choice of $\ell = O(\log{d})$.
By a union bound over all the blocks, we have
$\| \matu_{(b)} \|_F^2 \le \ell kR $
for all $b$ with probability of at least $1 - d^{-c}$.
Applying Lemma~\ref{lem:shortupper} and summing over these blocks gives:
\begin{align*}
\matashort^T \matashort \preceq \ell kR \cdot \mata^T \mata    
\end{align*}
Let $\matproj$ the projection operator onto the range space of $\matashort^T
\matashort$. Applying Lemma~\ref{lem:pseudoinversereverse} with $\matc = \matashort^T
\matashort$ and $\matd = \ell kR  \mata^T \mata  - \matashort^T \matashort$ gives
\begin{align*}
  \matproj \rbr{\mata^T
\mata}^\dag\matproj \preceq \ell kR \cdot \matproj\rbr{\matashort^T
\matashort}^\dag \matproj
\end{align*}

Further note that $\vecashort_i$ is completely contained within the range space of
$\matashort^T \matashort$.
Therefore for all $i$, $\matproj \vecashort_i = \vecashort_i$ and:
\begin{align*}
\str{\mata}{\vecashort_i}
& = \vecashort_i \matproj \rbr{\mata^T \mata}^\dag\matproj \vecashort_i^T\\
& \leq \ell kR \cdot \vecashort_i \matproj \rbr{\matashort^T \matashort}^\dag\matproj \vecashort_i^T\\
& = \ell kR \cdot \str{\matashort}{\vecashort_i}
\end{align*}
Therefore
\begin{align*}
\Pr \sbr{ \ell kR \cdot \str{\matashort}{\vecashort_i} \geq \str{\mata}{\vecashort_i}} 
& \ge \Pr \sbr{ \matproj \matashort^T \matashort \matproj \preceq \ell kR \cdot  \matproj \mata^T \mata  \matproj } \nonumber \\
& \ge 1 - d^{-c}.
\end{align*}
\QED

Combining Lemmas~\ref{lem:leverageupper}~and~\ref{lem:goodapprox}
shows that with high probability, scaling up $\str{\matashort}{\matashort_{(b)}}$
by $O(R^2 \log{d})$ gives upper bounds
for the leverages scores in the original blocks of $\mata$.

\begin{corollary}
\label{cor:projectiongood}
For any constant $c$, there exists a setting of constants
such that for any $R = d^{\theta}$, we have with probability
at least $1 - d^{-c}$
\begin{align*}
c' R^2 \log{d} \cdot \str{\matashort}{\matashort_{(b)}}
\geq \str{\mata}{\mata_{(b)}}
\end{align*}
holds for all $b$.
\end{corollary}


\subsection{Iterative Algorithm}

\begin{figure}[th!]
  \centering
  \includegraphics[width=12cm]{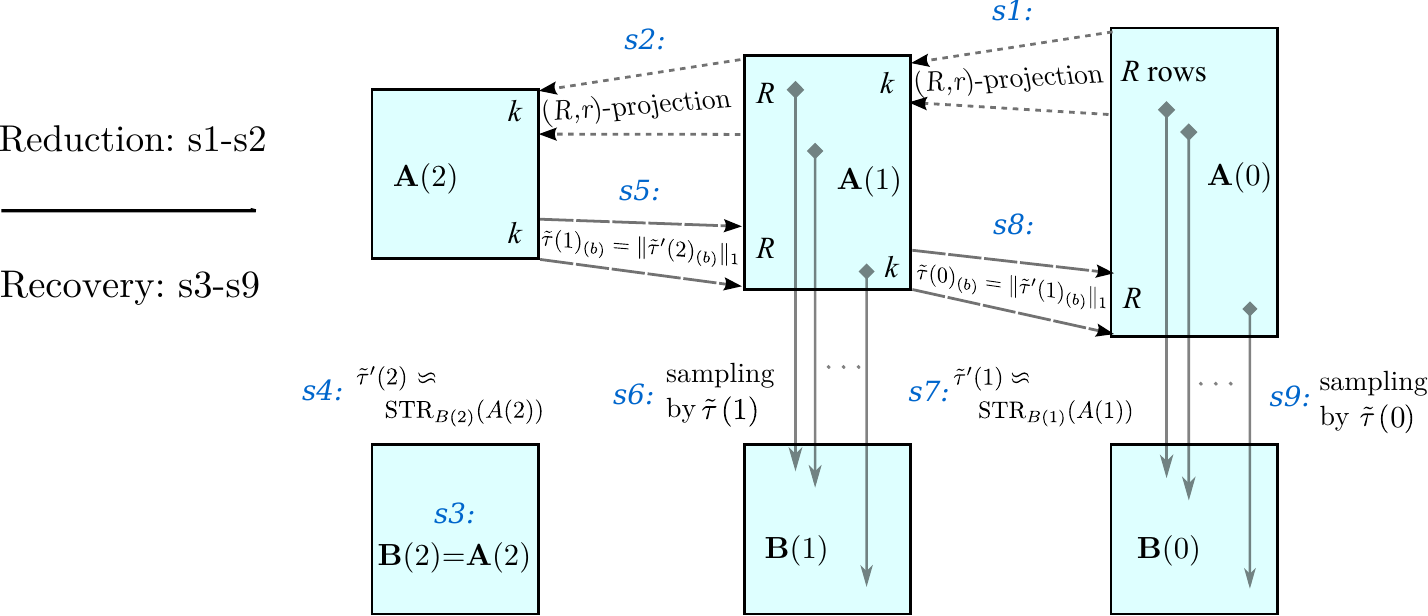}
  \caption{Illustration of Algorithm~\ref{alg:rowsamplel2} by using $L=2$. There are
    mainly two stages with 9 steps. On the reduction stage, we obtain shorter $\mata(1)$ and
    $\mata(2)$ by  iteratively doing $(R,r)$-projection. On the next recovery stage, we
    approximate the leverage scores of $\mata(1)$ by the ones computed from $\mata(2)$ and
    $\matb(2)$. Then $\matb(1)$ is sampled based on this approximated scores, which will be
  used to further obtain the approximated leverage scores of $A(0)$. The final step
  which samples $\matb(0)$ once again is not shown here.}
  \label{fig:rowsample}
\end{figure}

It remains to algorithmize the estimates that we obtain
using this projection process.
Projecting $\mata$ to $\matashort$ gives a matrix with
fewer rows, and a way to reduce the sizes of our problems.
A fast algorithm follows by examining the sequence of matrices
$\mata = \mata(0),  \mata(1), \ldots \mata(L)$ obtained
using such projections.
Once $\mata(L)$ has fewer than $\nnz(\mata) d^{-3}$ rows,
$\mata(L)^T \mata(L)$ can be approximated directly.
This then allows us to approximate the statistical leverage
scores of the rows of $\mata(L)$ 
Corollary \ref{cor:projectiongood} shows that stretches computed
on $\mata(l)$, $\approxleverage(l)$ can serve as sampling
probabilities in $\mata(l - 1)$.
This means we can gradually propagate solutions backwards
from $\mata(L)$ to $\mata(0)$.
We do so by maintaining the invariant that $\matb(l)$
has a small number of rows and is close to $\mata(l)$.
The total generalized stretch of $\mata(l)$ w.r.t. $\matb(l)$
can be used as upper bounds of the statistical leverage scores
of $\mata(l - 1)$ after suitable scaling..
This allows the sampling process to compute $\matb(l - 1)$,
keeping the invariant for $l - 1$.
Pseudocode of the algorithm is shown in Algorithm \ref{alg:rowsamplel2}, which is
illustrated in Figure~\ref{fig:rowsample}.

\begin{algo}[ht]
\qquad

$\textsc{RowSampleL2}(\mata, R, \epsilon)$
\vspace{0.05cm}

\underline{Input:}
Reduction rate $R$,
$n \times d$ matrix $\mata$,
allowed approximation error $\epsilon$,
failure probability $\delta = d^{-c}$.

\underline{Output:}
Sparsifier $\matb$ that contains
$O(R^{5} d \log{d} / \epsilon^{2})$
scaled rows of $\mata$
such that $(1 - \epsilon) \mata^T \mata
\preceq \matb^T \matb
\preceq (1 + \epsilon) \mata^T \mata$.

\begin{algorithmic}
\STATE{Set $L = \lceil \log_{R} (n / d) \rceil$}
\STATE{Set $\epsilon(0) = \epsilon / 3$, $\epsilon(1) \ldots \epsilon(l) = 1/2$}
\STATE{$\mata(0) = \mata$}
\FOR{$l = 1 \ldots L$}
        \STATE{Let $\mata(l)$ be a $(R, k)$-projection of $\mata(l - 1)$}
\ENDFOR
\STATE{$\matb(L) \leftarrow \mata(L)$} 
\FOR{$i = L \ldots 1$}
        \STATE{$\approxleveragev'(l) \leftarrow
                        O(R^3 \log{d}) \cdot \textsc{ApproxStr} (\mata(l), \sqrt{\frac{2}{3}}\matb(l), R, R)$}
        \STATE{Compute $\approxleverage(l - 1)$ by setting each entry in $\approxleverage(l - 1)_{(b)}$
                to $|\approxleveragev'(l)_{(b)}|_1$}
        \STATE{$\matb(l - 1) \leftarrow \textsc{Sample}(\mata(l - 1), \approxleverage(l - 1), \epsilon(l))$}
\ENDFOR
\STATE{$\approxleveragev'(0) \leftarrow
                        \textsc{ApproxStr} (\matb(0), \matb(0), 2, 2)$}
\RETURN{$\textsc{Sample}(\matb(0), \approxleveragev'(0), \epsilon / 3)$}

\end{algorithmic}
\caption{Row Sampling using Projections}

\label{alg:rowsamplel2}

\end{algo}

Sampling probabilities for $\mata(l - 1)$ are obtained by
computing the stretch of $\mata(l)$ w.r.t. $\matb(l)$.
We first show these values, $\approxleverage(l - 1)$, are
with high probability upper bounds for the statistical
leverage scores of $\mata(l - 1)$, $\leverage(l - 1)$.

\begin{lemma}
\label{lem:recovery}
Assume $\mata(l)$ and $\matb(l)$ satisfy  the following condition
\begin{align*}
        \frac{1}{2} \mata(l)^T \mata(l)\preceq
        \matb(l)^T \matb(l)
        \preceq \frac{3}{2} \mata(l)^T \mata(l)
      \end{align*}
Then for  any constant $c$, there is a setting of the constants
such that 
\begin{itemize}
        \item $\approxleveragev(l - 1) \geq \leveragev(l - 1)$
        \item $\| \approxleveragev(l - 1) \|_1 \leq O(dR^3 \log{d})$
        \end{itemize}
        holds with 
 probability at least $1 - d^{-c}$.
\end{lemma}

\Proof
The given condition implies that $\sqrt{\frac{2}{3}}\matb(l)$
satisfies the condition needed for Lemma \ref{lem:approxstretch}
with $\kappa = 3$. Let the constants $c'=c+\log_d 2$, then with probability
at least $1 - d^{-c'}$ we have:
\begin{align*}
\approxleveragev'(l) \geq O(R^2\log d) \leveragev(l)
\end{align*}
Since $\mata(l)$ is a projection of $\mata(l - 1)$, we
can index corresponding blocks in them.
Apply Corollary \ref{cor:projectiongood}, then with probability at least $1-d^{-c'}$, we have
\begin{align*}
  O(R^2 \log d) \nbr{\leveragev(l)_{(b)}}_1 \ge \nbr{\leveragev(l-1)_{(b)}}_1
\end{align*} for all blocks. As each entries of $\approxleveragev(l-1)$ in block $b$ is
assigned with $\nbr{\approxleveragev '(l)_{(b)}}_1$, by the union bound, then for all $i$ in block $b$ 
\begin{align*}
  \approxleveragev(l-1)_i = \nbr{\approxleveragev'(l)_{(b)}}_1 \ge \nbr{\leveragev(l-1)_{(b)}}_1  \ge \leveragev(l-1)_{i}
\end{align*}
holds for all blocks $b$ with probability at least $1-2d^{-c'}$, which is equal to $1-d^{-c}$ by the
definition of $c'$.

It remains to upper bound $\nbr{\approxleveragev(l - 1)}_1$.
Lemma \ref{lem:approxstretch} Part \ref{part:l2stretchsum} gives,
with probability at least $1 - d^{-c}$,
$\nbr{\approxleveragev'(l)}_1 \leq O(dR^2\log d)$.
As each $\nbr{\approxleveragev'(l)_{(b)}}_1$ is assigned to the $R$
entries in $\approxleveragev(l - 1)_{(b)}$, we get
$\nbr{\approxleveragev(l - 1) }_1 \leq  O ( d R^{3} \log{d} )$ with the same probability. 
\QED

Combining these with the fact that the number of rows decrease
by a factor of $O(R)$ per iteration completes the algorithm.
Our main result for $\ell_2$ row sampling is
obtained by setting $R$ to $d^{\theta}$.
Applying Lemma~\ref{lem:recovery} inductively backwards in $l$
gives the overall bound.

\begin{theorem}
\label{thm:algol2}
For any constant $c$, there is a setting of constants
such that if $\textsc{RowSampleL2}$, shown in Algorithm~\ref{alg:rowsamplel2}, is ran with
$R = d^{\theta}$, then with probability at least $1 - d^{-c}$ it returns $\matb$
in $O(\nnz(\mata) + d^{\omega + 4 \theta} \epsilon^{-2})$ time
 such that:
\begin{align*}
(1 - \epsilon) \mata^T \mata
\preceq \matb^T \matb
\preceq (1 + \epsilon) \mata^T \mata
\end{align*}
and $\matb$ has
${O}(d \log{d} \epsilon^{-2})$
rows, each being a scaled copy of some row of $\mata$,
\end{theorem}

\Proof
We first show correctness via.~induction backwards on $l$.
Define $c'=c+1$, we show that $\matb(l)$
has $O(d R^{4} \log{d}\epsilon(l)^{-2})$ rows
and satisfies $$(1 - \epsilon(l)) \mata(l)^T \mata(l)^T
\preceq \matb(l)^T \matb(l)
\preceq (1 + \epsilon(l)) \mata(l)^T \mata(l)^T,$$
with probability at least
$1 - 3(L - l) d^{c'}$ for each $l$.
 
As $k=O({c}/{\theta})$ is a constant, one $(R,k)$-projection decreases the number of
rows by a factor of $O(R)$. After $L=\log_{R}(n / d)$ projections, we get that $\mata(L)$
has $O(d)$ rows.
Therefore the base case where $l = L$ follows from
$\matb(L) = \mata(L)$.

For the inductive step, we assume that the inductive hypothesis
holds for $l \geq 1$ and try to show it for $l - 1$.
As $\epsilon(l)$ was set to $1/2$, we have:
\begin{align*}
        \frac{1}{2} \mata(l)^T \mata(l)\preceq
        \matb(l)^T \matb(l)
        \preceq \frac{3}{2} \mata(l)^T \mata(l)
\end{align*}
This allows us to invoke Lemma \ref{lem:recovery},
which combined with Lemma \ref{lem:samplel2}
gives that with probability $1 - d^{-c' }$,
the inductive hypothesis also holds for $l - 1$.

The final sampling step on $\matb(0)$ with $R=2$ guarantees that
$\nbr{\approxleveragev'(0)}_1 \le O(d)$, which gives a final row
count of $O(d \log d/\epsilon^2)$.
The overall failure probability follows from union bounding this
with the failure probabilities from the hypothesis
and Lemma \ref{lem:recovery}.

We now bound the total running time, starting with the projections. Since $k=O(1)$, each
$(R,k)$-projection reduces $R$ rows into $O(1)$ rows, so the sparsity patterns are kept,
namely $\nnz(\mata(l + 1)) = O(\nnz(\mata(l)))$. As  $\matu_{(b)}$ has $k$ rows for all
$b$, then the cost of constructing $\mata(l)$ from $\mata(l-1)$ is $O(\nnz(\mata))$. 
Note that since $R = d^{\theta}$,
$L = \log_{R}(n / d) = \log_R(\poly(d))$ is a constant.
So we get a total cost of $O(\nnz(\mata))$ over all $L$ projections.

Since $\matb(l)$ has $O(d R^{3} \log^2 d  \epsilon(l)^{-2})$ rows,
Lemma \ref{lem:approxstretch} gives that each call to $\textsc{ApproxStr}$ takes
$O(\nnz(\mata) + d^{\omega+4\theta}  \epsilon^{-2})$ time, where we upper bound
$\log^2{d}$ by $d^\theta$. The cost of the final step on $\matb(0)$ can be
bounded similarly.
\QED

%% file: pnorm.tex
\section{Algorithm for Preserving $\ell_p$-norm}
\label{sec:pnorm}

We now turn to the more general problem of finding $\matb$
with $\poly(d)$ rows such that:
\begin{align*}
(1 - \epsilon) \nbr{\mata \vecx}_p
\leq \nbr{\matb \vecx}_p
\leq (1 + \epsilon) \nbr{\mata \vecx}_p
\qquad \forall \vecx \in \Re^{d}
\end{align*}
We will make repeated use of the following (tight) inequality
between $\ell_2$ and $\ell_p$ norms, which can be obtained
by direct applications of power-mean and H\"{o}lder's inequalities.

\begin{fact}
\label{fact:holdersimple}
Let $\vecx$ be any vector in $\Re^{d}$, and $p$ and $q$
any two norms where $1 \leq p \leq q$, we have:
\begin{align*}
\nbr{\vecx}_q \le \nbr{\vecx}_p
 & \le  d^{\frac{1}{q}-\frac{1}{p}}\nbr{\vecx}_q
\end{align*}
\end{fact}
We will use this Fact with one of $p$ or $q$ being $2$,
in which case it gives:
\begin{itemize}
        \item If $1 \leq p \leq 2$, $\nbr{\vecx}_2 \le \nbr{\vecx}_p \le d^{\frac{1}{2}-\frac{1}{p}}\nbr{\vecx}_2$
        \item If $2 \leq p$, $d^{\frac{1}{p} - \frac{1}{2}} \nbr{\vecx}_2 \le \nbr{\vecx}_p \le \nbr{\vecx}_2$
\end{itemize}

As $\mata \vecx \in \Re^{n}$, its $\ell_2$ and $\ell_p$ norms can differ
by a factor of $\poly(n)$.
This means the $\ell_2$ row sampling algorithm from 
Section~\ref{sec:algo2} can lead to $\poly(n)$ distortion.
Our algorithm in this section can be viewed as a way to reduce
this distortion via. a series of iterative steps.
Once again, our algorithm is built around a sampling concentration bound.
The sampling probabilities are based on the definition of a well-conditioned
basis, which is more flexible than $\ell_2$ statistical leverage scores.

\begin{definition}
\label{def:wellcondbasis}

Let $\mata$ be an $n \times d$ matrix of rank $r$, $p \in [1, \infty]$ and
$q$ be its dual norm such that $\frac{1}{p} + \frac{1}{q} = 1$.
Then an $n \times r$ matrix $\matu$ is an $(\alpha, \beta, p)$-well-conditioned
basis for the column space of $\mata$ if the columns of $\matu$ span the
column space of $\mata$ and:
\begin{enumerate}
        \item $\vertiii{\matu}_p \leq \alpha$.
        \item For all $\vecx \in \Re^{r}$, $\nbr{\vecx}_q
                \leq \beta \nbr{\matu \vecx}_p$.
\end{enumerate}

\end{definition}

A $\ell_p$ analog of the sampling concentration result given in
Lemma \ref{lem:samplel2} was shown in \cite{DasguptaDHKM09}.
It can be viewed a generalization of Lemma~\ref{lem:samplel2}.

\begin{lemma}
\label{lem:samplelp}
(Theorem 6 of \cite{DasguptaDHKM09})
Let $\mata$ be a $n \times d$ matrix with rank $r$,
$\epsilon \leq 1/7$, and let $p \in [1, \infty)$.
Let $\matu$ be an $(\alpha, \beta, p)$-well-conditioned basis for $\mata$.
Then for any sampling probabilities $\vecp \in \Re^{n}$ such that:
\begin{align*}
p_i & \geq c_p (\alpha \beta)^{p} \frac{\nbr{\matu_{i*}}^p_p}{\vertiii{\matu}^p_p},
\end{align*}
where $\matu_{i*}$ is the $i$-th row of $\matu$ and $c_p$ is a constant  depending only on
$p$. Then with  probability at least $1 - d^{-c}$,
$\textsc{Sample}(\mata, p, \epsilon)$ returns $\matb$ satisfying 
\begin{align*}
 (1-\epsilon)  \nbr{\mata \vecx}_p  \le  \nbr{\matb \vecx}_p \leq (1+\epsilon) \nbr{\mata \vecx}_p
                \qquad \forall \vecx \in \Re^{d}
\end{align*}
\end{lemma}

We omitted the reductions of probabilities that are more than $1$ since
this step is included in our formulation of \textsc{Sample}.
Several additional steps are needed to turn this into an
algorithmic routine, the first being computing $\matu$.
A na\"{i}ve approach for this requires matrix multiplication, and the
size of the outcome may be more than $\nnz(\mata)$.
Alternatively, we can find a linear transform used to create it,
aka. a matrix $\matc$ such that $\matu = \mata \matc$.

The estimation of $\nbr{(\mata \matq)_{i*}}_p^p$ and sampling
can then be done in a way similar to Section~\ref{sec:algo2}.
When $1 \leq p \leq 2$, we can compute $O(1)$ approximations
using $p$-stable distributions \cite{Indyk06} in a way analogous
to Section 4.2.1. of Clarkson et al. \cite{ClarksonDMMMW12}.
When $2 \leq p$, we will use the $2$-norm as a surrogate at
the cost of more rows.
As all of our calls to \textsc{Sample} will be using probabilities
estimated via. the same matrix, we will the estimation
of $p$-norm leverage scores and sampling as a single blackbox.

\begin{lemma}
\label{lem:estimateandsamplelp}
For any constant $c$, there exist an algorithm
$\textsc{EstimateAndSampleP}(\mata, \matc, \alpha, \beta, R, \epsilon)$
that given a $\mata$, $\matc$ such that $\mata \matc$ is a
$(\alpha, \beta, p)$-well-conditioned basis for $\mata$,
returns a matrix $\matb$ with probability at least $1 - d^{-c}$  such
that:
\begin{equation*}
  (1-\epsilon)  \|\mata \vecx\|_p \le \|\matb \vecx\|_p  \le (1+\epsilon) \|\mata \vecx\|_p 
\end{equation*}
in $O(\nnz(\mata) \log_{R}(d)  + R d^{\omega} \log{d})$ time and
the number of rows in $\matb$ can be bounded by:
\begin{equation*}
\begin{cases}
O((\alpha \beta R)^{p} d \log(d) \epsilon^{-2}) & \textrm{if }1 \leq p \leq 2 \\
O((\alpha \beta R)^{p} d^{\frac{p}{2}} \log(d) \epsilon^{-2}) & \textrm{if }2 \leq p
\end{cases}
\end{equation*}
\end{lemma}

\subsection{Sampling Using $\ell_2$-leverage scores}
\label{subsec:oneshot}

Our starting point is the observation that a good basis for $\ell_2$,
specifically a nearly orthonormal basis of $\mata$
still allows us to reduce number of rows substantially under $\ell_p$.

\begin{lemma}
\label{lem:simplebasis}
If $\matc \in \Re^{d \times r}$ satisfies
$\frac{1}{2} (\mata^T \mata)^{\dag}
\preceq \matc \matc^T \preceq \frac{3}{2} (\mata^T \mata)^{\dag} $
then $\matu = \mata \matc$ is a $(\alpha, \beta, p)$-well-conditioned basis
for $\mata$ where
$\alpha \beta \leq O(n^{|\frac{1}{2} - \frac{1}{p}|} d^{|\frac{1}{2} - \frac{1}{p}| + \frac{1}{2}} )$.
\end{lemma}

\Proof
We start by show that $\matu^T \matu$ is close to the identity matrix
as an operator.
Also, since $\matc^T \matc$ is a full rank matrix and
$\matc^T (\matc \matc^T)^{\dag} \matc$ is a projection operator
onto the column space of $\matc$, we have
$\matc^T (\matc \matc^T)^{\dag} \matc = \mati.$
Taking pseudoinverses of the given condition on $\matc$ gives:
\begin{align*}
\frac{2}{3} (\matc \matc^T)^{\dag}
& \preceq \mata^T \mata \preceq 2 (\matc \matc^T)^{\dag}
\end{align*}
Substituting it into $\matu = \mata \matc$ then gives:
\begin{align*}
\matu^T \matu
 = \matc^T \mata^T \mata \matc
 \preceq 2\matc^T (\matc \matc^T)^{\dag} \matc
 = 2\mati
\end{align*}
 and
\begin{align*}
  \frac{2}{3}\mati  \preceq  \frac{2}{3}\matc^T (\matc \matc^T)^{\dag} \matc \preceq
 \matu^T \matu
\end{align*}

This allows us to infer that $\vertiii{\matu}_2 \leq  \sqrt{2d}$,
and for any vector $\vecx$, $\nbr{\vecx }_2 \leq \sqrt 2 \nbr{\matu \vecx}_2$.

Next we find values of $\alpha$ and $\beta$ that meet the
requirements of a well-conditioned basis given in
Definition~\ref{def:wellcondbasis}. Let $q$ be the dual norm for $p$ which
satisfies $\frac{1}{p}+\frac{1}{q}=1$. 

First consider the case where $1 \leq p \leq 2$.
We can view all entries of the matrix $\matu$ as a vector
of length $nr \leq nd$ vector.
Apply Fact \ref{fact:holdersimple} gives:
\begin{align*}
\vertiii{\matu}_p
& \le (nd)^{\frac{1}{p}-\frac{1}{2}}\vertiii{\matu}_2
\end{align*}

Which gives $\vertiii{\matu}_2 \le \sqrt{2} d^{\frac{1}{2}}$ and therefore
$\alpha = \sqrt{2} (nd)^{\frac{1}{p}-\frac{1}{2}}d^{\frac{1}{2}}$.
For the second part, given any vector $\vecx$, we have
\begin{align*}
\nbr{\vecx}_q
& \leq  \nbr{\vecx}_2
        && \text{(by Fact~\ref{fact:holdersimple} on $\vecx$ since $q \leq 2$)}\\
& \leq \sqrt 2\nbr{\matu \vecx}_2\\
& \leq \sqrt 2\nbr{\matu \vecx}_p
        && \text{(by Fact~\ref{fact:holdersimple} on $\matu \vecx$ since $1 \leq p \leq 2$)}\\
\end{align*}
which means $\beta = \sqrt 2$ suffices.

Now we consider the case where $p \geq 2$ similarly.
\begin{align*}
\vertiii{\matu}_p \le \vertiii{\matu}_2 \le \sqrt{2} d^{1/2}
\end{align*}
and:
\begin{align*}
\nbr{\vecx}_q
& \le d^{\frac{1}{q}-\frac{1}{2}}\nbr{\vecx}_2
        && \text{(by Fact~\ref{fact:holdersimple} on $\vecx$ since $q \leq 2$)}\\
&= d^{\frac{1}{2} - \frac{1}{p}}\nbr{\vecx}_2 
        && \text{(since $\frac{1}{q} = 1 - \frac{1}{p}$)}\\
& \leq \sqrt 2 d^{\frac{1}{2}-\frac{1}{p}}\nbr{\matu \vecx}_2\\
& \le \sqrt 2 d^{\frac{1}{q}-\frac{1}{2}}n^{\frac{1}{2}-\frac{1}{p}}\nbr{\matu \vecx}_p
        && \text{(by Fact~\ref{fact:holdersimple} on $\matu \vecx$ with $2 \leq p$)}\\
& = \sqrt 2 (nd)^{\frac{1}{2}-\frac{1}{p}}\nbr{\matu \vecx}_p\
\end{align*}

Combining the bounds from these two cases on $p$ gives that
$U$ is a $(\alpha,\beta,p)-$well-conditioned basis, where
  \begin{equation*}
  \alpha =
  \begin{cases}
    \sqrt{2} d^{\frac{1}{2}} & \textrm{for } p \ge 2\\
    \sqrt{2} (nd)^{\frac{1}{p}-\frac{1}{2}}d^{\frac{1}{2}} & \textrm{otherwise } 
\end{cases}
\quad \textrm{and} \quad 
\beta =
\begin{cases}
  \sqrt 2 (nd)^{\frac{1}{2}-\frac{1}{p}}   & \textrm{for } p \ge 2\\
  \sqrt 2 & \textrm{otherwise}
\end{cases}
  \end{equation*}
It can be checked that in both cases the stated bound on
$\alpha \beta$ holds.
\QED

One way to generate such a nearly-orthonormal basis is by the
$L_2$ approximation that we computed in Section~\ref{sec:algo2}.
This leads to a fast algorithm, but the dependency on $n$ in this
bound precludes a single application of sampling using the values given because each time
we transfer $n$ rows into $O(p|\frac{1}{2}-\frac{1}{p}|)$ rows. 
However, note that when $p < 4$,
$p |\frac{1}{2} - \frac{1}{p}| = |1 - \frac{p}{2}| < 1$.
This means it can be used as a reduction step in an iterative algorithm
where the number of rows will decrease geometrically.
Therefore, we can use this process as a reduction routine.
For inductive purposes, we will also state the routine to compute
the basis via. an approximation of $\mata$, $\mataapprox$.
Pseudocode of our reduction algorithm is given in Algorithm~\ref{alg:reductionlp}.

\begin{algo}[ht]
\qquad

$\textsc{ReduceP}(\mata, \mataapprox, \epsilon)$
\vspace{0.05cm}

\underline{Input:}
$n \times d$ matrix $\mata$ and its approximated 
matrix  $\mataapprox$,  $p$-norm and error parameter $\epsilon$

\underline{Output:}
Matrix $\matb$

\begin{algorithmic}[1]
  \STATE{$\matashort \gets \textsc{RowCombine}(\mataapprox, p, 1/3, d^{-c - 1})$}
  \STATE{Perform SVD on $\matashort^T \matashort$ and then construct $\matc$ by dropping
    the zero singular values and corresponding singular vectors such that $\matc^T\matc = (\matashort^T \matashort)^{\dag}$ }
  \IF{$1 \leq  p \leq 2$}
        \STATE{$\alpha \leftarrow     \sqrt{2} (nd)^{\frac{1}{p}-\frac{1}{2}}d^{\frac{1}{2}}$,
                $\beta \leftarrow \sqrt{2}$}
  \ELSE
        \STATE{$\alpha \leftarrow \sqrt{2} d^{\frac{1}{2}}$,
                $\beta \leftarrow \sqrt 2 (nd)^{\frac{1}{2}-\frac{1}{p}}$}
  \ENDIF
  \STATE{$\matb \leftarrow \textsc{EstimeAndSampleP}(\mata, \matc, 2 \alpha, 2 \beta, p, d^{\frac{\theta}{2p}}, \epsilon)$}
  \RETURN{$\matb$}
\end{algorithmic}

\caption{Reduction Step for Preserving $\ell_p$ Norm}

\label{alg:reductionlp}
\end{algo}

\begin{lemma}
\label{lem:reductionlp}
For any constant $c$, there exist a setting of constants
in $\textsc{ReduceP}$ such that if $\mata$ and $\mataapprox$ satisfy
\begin{equation*}
\frac{1}{2} \nbr{\mata \vecx}_p \le \nbr{\mataapprox \vecx}_p
        \leq \frac{3}{2} \nbr{\mata \vecx}_p \qquad \forall \vecx \in \Re^{d}
\end{equation*}
and $\mataapprox$ has $\tilde{n}$ rows, then
$\textsc{ReduceP}(\mata, \mataapprox, \epsilon)$
returns in $O(\nnz(\mata) \log{d+\theta} + d^{\omega+\theta} \log{d})$ time
a matrix $\matb$ such that with probability at least $1 - d^{-c}$:
\begin{equation*}
(1-\epsilon) \nbr{\mata \vecx}_p \le \nbr{\matb \vecx}_p
        \leq (1+\epsilon) \nbr{\mata \vecx}_p \qquad \forall \vecx \in \Re^{d}
\end{equation*}
And the number of rows in $\matb$ can be bounded by:
\begin{align*}
\begin{cases}
O(n^{1 - \frac{p}{2}} d^{2 + \theta} \epsilon^{-2}) &\text{if } 1 \leq p \leq 2\\
O(n^{\frac{p}{2} - 1} d^{\frac{3}{2}p - 1 + \theta} \epsilon^{-2}) &\text{if } 2 \leq p
\end{cases}
\end{align*}
\end{lemma}

The proof will be in two steps: we first show that $\mataapprox \matc$ is a
well-conditioned basis for $\mataapprox$, and use the following Lemma
to show that this implies that $\mata \matc$ is a well-conditioned
basis for $\mata$.

\begin{lemma}
\label{lem:basereverse}
If $\mata$ and $\mataapprox$ are such that for all $\vecx \in \Re^{d}$,
$ \frac{1}{2}\nbr{\mata \vecx}_p  \le \|\mataapprox \vecx\|_p  \leq \frac{3}{2} \nbr{\mata \vecx}_p$,
and $\matc$ is such that $\matuapprox = \mataapprox \matc$ is an
$(\alpha, \beta, p)$-well-conditioned basis for $\mata$, then
$\mata \matc$ is also an $(2 \alpha, 2 \beta, p)$-well-conditioned
basis for $\mata$.
\end{lemma}

\Proof
It suffices to verify both conditions of Definition~\ref{def:wellcondbasis}
holds for $\matu = \mata \matc$.
For $\vertiii{\matu}_p$, we can treat its $p$\textsuperscript{th}
power as a summation over the columns of $\matu$ and get:
\begin{align*}
\vertiii{\matu}_p^p
& = \sum_{j = 1}^d \nbr{\matu_{*j}}_p^p
= \sum_{j = 1}^d \nbr{\mata \matc_{*j}}_p^p \\
& \leq \sum_{j = 1}^d {2}^p \nbr{\tilde \mata \matc_{*j}}_p^p && \text{(by
  assumption $\frac{1}{2}\nbr{\mata \vecx}_p  \le \|\mataapprox \vecx\|_p$)}\\
& = {2}^p \vertiii{\matuapprox}^p_p = 2^p \alpha^p
\end{align*}
The other condition can be obtained by direct substitution.
By the condition given, we have that for all $\vecz \in \Re^{d}$:
\begin{align*}
\nbr{\vecz}_q
& \leq \beta \nbr{\matuapprox \vecz}_p
= \beta \nbr{\mataapprox \matc \vecz}_p
\end{align*}
Applying the fact that $\nbr{\mata \vecx}_p \leq 2 \|{\mataapprox \vecx}\|_p$
to the vector $\matc \vecz$ gives:
\begin{align*}
\nbr{\vecz}_q
& \leq \beta (2 \nbr{\mata \matc \vecz}_p)
= 2 \beta \nbr{\matu \vecz}_p
\end{align*}
\QED

\Proofof{Lemma~\ref{lem:reductionlp}}
By the guarantees of \textsc{RowCombineL2} given in Theorem~\ref{thm:algol2},
we can set its constants so that with probability at least $1 - d^{-c'}$ we have:
\begin{align*}
\frac{1}{2} \mataapprox^T \mataapprox
\preceq \matashort^T \matashort
\preceq \frac{3}{2} \mataapprox^T \mataapprox.
\end{align*}
As $\matc^T \matc = (\matashort^T \matashort)^\dag$, then the condition of
Lemma~\ref{lem:simplebasis} is satisfied, so $\tilde \mata \matc$ is a well-conditioned
basis of $\tilde \mata$. Furthermore, by Lemma~\ref{lem:basereverse}, $\mata \matc$ is also a
well-conditioned basis of $\mata$.
The guarantees for $\matb$ then follows from Lemma~\ref{lem:estimateandsamplelp}. The
probability can be obtained by a simply union bound with $c'=c+\log 2$.
\QED



Iterating this reduction routine with $\mataapprox = \mata$
gives a way to reduce the row count
from $n$ to $\poly(d)$ in $O(\log\log(n / d))$ iterations when $p < 4$.
Two issues remain: the approximation errors will accumulate across
the iterations, and it's rather difficult (although possible if
additional factors of $d$ are lost) to bound the reductions of non-zeros
since different rows may have different numbers of them.
We will address these two issues systematically before giving
our complete algorithm.

The only situation where a large decrease in the number of rows
does not significantly decrease the overall number of non-zeros
is when most of the non-zeros are in a few rows.
A simple way to get around this is to `bucket' the rows of $\mata$
by their number of non-zeros, and compute $\poly(d)$ sized
samples of each bucket separately.
This incurs an extra factor of $\log{d}$ in the final number of
rows, but ensures a geometric reduction in problem sizes
as we iterate.

The error buildup can in turn be addressed by sampling on the rows of
the initial $\mata$ using the latest approximation for it $\mataapprox$.
However, since the algorithm can take up to $O(\log{d})$ iterations,
we need to perform this on a reduced version of $\mata$ instead to
obtain a $O(\nnz(\mata))$ running time.
Pseudocode of our algorithm for a single partition where the number of
non-zeros in each row are within a constant factor of each other
is given in Algorithm~\ref{alg:rowsamplelp}.

\begin{algo}[ht]
\caption{Algorithm for Producing Row Sample of Size $\poly(d)$ that Preserves $\ell_p$-norm}
\label{alg:rowsamplelp}

\qquad

$\textsc{RowSampleP}(\mata, p, \epsilon, \delta)$
\vspace{0.05cm}

\underline{Input:}
$n \times d$ matrix $\mata$, $p$, error parameter $\epsilon$, failure probability $\delta = d^{-c}$

\underline{Output:}
Matrix $\matb$

\begin{algorithmic}[1]
  \IF{$1 \leq p \leq 2$}
        \STATE{$n^{*} \leftarrow O(d^{\frac{4}{p} + \theta}
                \log^{\frac{2}{p}}d)$} 
  \ELSE
        \STATE{$n^{*} \leftarrow O(d^{\frac{3}{2}p - 1 + \theta}
                \log^{\frac{3p - 2}{4 - p}}{d})$} 
  \ENDIF
  \IF{$\mata$ has $n^{*}$ or fewer rows}
      \RETURN{$\mata$}
  \ENDIF
  \STATE{$\mataapprox_0, \mataapprox \leftarrow \textsc{ReduceP}(\mata, \mata, 1/5)$}
  \WHILE{$\mataapprox$ has more than $\bar{n}$ rows}
        \STATE{$\mataapprox \leftarrow \textsc{ReduceP}(\mataapprox_0, \mataapprox, 1/5)$}
  \ENDWHILE
  \STATE{$\matb \leftarrow \textsc{ReduceP}(\mata, \mataapprox, \epsilon / 2)$}
  \RETURN{$\matb$}
\end{algorithmic}

\end{algo}

\begin{lemma}
\label{lem:rowsamplelp}
For any $c$, there is a setting of constants in \textsc{RowSample} such that
given a matrix $\mata$ where each row has between $[s, 2s]$ nonzeros and $p < 4$.
$\textsc{RowSampleP}(\mata, p, \epsilon)$
with probability $1 - d^{-c}$ returns
in $O(\nnz(\mata) + d^{\omega} \log{d})$ time a matrix $\matb$
such that:
\begin{align*}
(1-\epsilon)\nbr{\mata \vecx}_p \le  \nbr{\matb \vecx}_p 
        \leq (1+\epsilon) \nbr{\mata \vecx}_p 
\end{align*}
And the number of rows in $\matb$ can be bounded by 
\begin{align*}
\begin{cases}
O(d^{\frac{4}{p} + \theta } \epsilon^{-2})
        &\text{if } 1 \leq p \leq 2\\
O(d^{\frac{3}{2}p - 1 + \theta }  \epsilon^{-2}) &\text{if } 2 \leq p \leq 4
\end{cases}
\end{align*}
\end{lemma}

\Proof
For correctness, we can show by induction that as long as
all calls to $\textsc{ReduceP}$ succeeds,
$\left| \nbr{\mataapprox_0 \vecx}_p - \nbr{\mataapprox \vecx}_p \right|
\leq 1/5 \nbr{\mataapprox_0 \vecx}_p$ for all $\vecx \in \Re^{d}$.
This can be combined with the guarantee between
$\mata$ and $\mataapprox$ to give:
$\left| \nbr{\mata \vecx}_p - \nbr{\mataapprox \vecx}_p \right|
\leq 1/2 \nbr{\mata \vecx}_p$ for all $\vecx \in \Re^{d}$,
which allows us to obtain the bound on $\matb$.
The analysis below shows that there can be at most $O(\log{d})$
calls to $\textsc{ReduceP}$, so calling each with success probability
at least $1 - d^{-c - 2}$ gives an overall success probability of
at least $1 - d^{-c - 1}$.

To bound the runtime, we first show that if $\matb$ has $n_b$ rows, then
in the next iteration the number of rows in $\matb$
can be bounded by $(n_b / n*)^{c_p} n*$
where $c_p < 1$ is a constant based on $p$.
When $1 \leq p \leq 2$, Lemma~\ref{lem:reductionlp} gives that
there exist some constant $c_0$ such that the new row count can be bounded by:
\begin{align*}
c_0 n_b^{1 - \frac{p}{2} }d^{2 + \theta} \log(d)
= (n_b c_0^{-\frac{2}{p}} d^{-\frac{4}{p} - \frac{2 \theta}{p}} \log^{-\frac{2}{p}} {d})^{1 - \frac{p}{2}}  c_0^{\frac{2}{p}} d^{\frac{4}{p} + \frac{2 \theta}{p}} \log^{\frac{2}{p}}{d}
\end{align*}
So $n^{*} = c_0^{\frac{2}{p}} d^{\frac{4}{p} + O(\theta)} \log^{\frac{2}{p}}{d}$
and $c_p = 1 - \frac{p}{2}$ suffices.
Similarly for the case where $2 \leq p$, it can be checked that
\begin{align*}
n^{*} = c_0^{\frac{2}{4 - p}} d^{\frac{3p - 2}{4 - p} + O(\theta)} \log^{\frac{2}{4 - p}}{d}
\end{align*}
Gives that the new row count can be bounded by $(n_b / n^{*})^{c_p} n^{*}$
where $c_p = \frac{p}{2} - 1$.
As we can set $\theta$ to any arbitrary constant, the constant
in front of its exponent can also be removed.

Therefore in $O(\log_{c_p^{-1}}(\log(n / n^{*})) = O(\log\log{n})$ iterations
the number of rows in $\matb$ decreases below $2 n^{*}$.
Also, the number of rows in $\mataapprox$ is at most
$(n / n^{*})^{c_p} n^{*} = n (n / n^{*})^{c_p - 1}$.
This means the total cost to obtain the final $\matb$
can be bounded by $O(\nnz(\mata) +
(\nnz(\mata) (n / n^{*})^{c_p - 1}  + d^{\omega + \theta} \log(n / n^{*})))$
Since $\log{t} \leq O(t^{1 - c_p})$, the first two terms can be
bounded by $O(\nnz(\mata))$.
The overall runtime bound then follows by applying Lemma~\ref{lem:reductionlp}
to the final call of $\textsc{ReduceP}$.


%
\QED


\subsection{Fewer Rows by Iterating Again}
\label{subsec:again}

A closer look at the proof of Lemma~\ref{lem:rowsamplelp}
shows that a significant increase in the number of rows comes
from dividing by the $1 - |1 - \frac{p}{2}|$ term in the
exponent of $n$.
As a result, the row count can be further reduced if the leverage
scores are computed via. a $p'$-norm approximation where $q$
is between $2$ and $p$.
For simplicity we only show this improvement for 
the case where $1 \leq p \leq p' \leq 2$.

We will start by proving a generalization of Lemma~\ref{lem:simplebasis}.
\begin{lemma}
\label{lem:basisq}
If $\mata$ has rank $r$, $1 \leq p \leq p' \leq 2$,
and $\mataapprox$ is a matrix with $\tilde{n}$ rows such that
for all vectors $\vecx$ we have
$\frac{1}{2}\nbr{\mata \vecx}_q \le \nbr{\mataapprox \vecx}_q 
\leq \frac{3}{2} \nbr{\mata\vecx}_q$,
and $\matc$ is a $d \times r$ matrix such that
$\frac{1}{2} (\mataapprox^T \mataapprox)^{\dag}
\preceq \matc \matc^T
\preceq \frac{3}{2} (\mataapprox^T \mataapprox)^{\dag}$,
then $\matu = \mata \matc$ is a $(\alpha, \beta, p)$-well-conditioned
basis for $\mata$ where $\alpha \beta \leq
O(n^{\frac{1}{p} - \frac{1}{p'}} \tilde{n}^{\frac{1}{p'} - \frac{1}{2}}
d^{\frac{1}{p}} )$.
\end{lemma}

\Proof
Let $\tilde{\matu} = \mataapprox \matc$.
Similar to the proof of Lemma~\ref{lem:simplebasis}, we have
$\vertiii{\tilde{\matu}}_2 \leq \sqrt{2 d}$,
and $\nbr{\vecx }_2 \leq \sqrt{2} \nbr{\tilde{\matu} \vecx}_2$ for any vector $\vecx$.
Once again, let $q$ be the dual norm for $p$ such that $\frac{1}{p}+\frac{1}{q}=1$. 

We have:
\begin{align*}
\vertiii{\mata \matc}_p
& \le (nd)^{\frac{1}{p}-\frac{1}{p'}}\vertiii{\mata \matc}_{p'} \\
& \le (nd)^{\frac{1}{p}-\frac{1}{p'}} \left( 3/2 \vertiii{\mataapprox \matc}_{p'} \right)\\
& \le 3/2 (nd)^{\frac{1}{p}-\frac{1}{p'}}
        (\tilde{n}d)^{\frac{1}{p'}-\frac{1}{2}}\vertiii{\mataapprox \matc}_2
\end{align*}
Which gives $\alpha = O(n^{\frac{1}{p} - \frac{1}{p'}}
        \tilde{n}^{\frac{1}{p'} - \frac{1}{2}} d^{\frac{1}{p}})$.
Also,
\begin{align*}
\nbr{\vecx}_q
 \leq  \nbr{\vecx}_2
 \leq 2\nbr{\mataapprox \matc \vecx}_{2}
 \leq 3\nbr{\mata \matc \vecx}_{p'}
 \leq 3 \nbr{\mata \matc \vecx}_{p}
\end{align*}
Which gives $\beta = O(1)$.

\QED

This allows us to compute leverage scores via. a $\ell_{p'}$-norm approximation.
By Lemma~\ref{lem:rowsamplelp}, such a matrix has
$\tilde{n} = O(d^{\frac{4}{p'}} \log^{\frac{2}{p'}}{d})$ rows.
By Lemma~\ref{lem:estimateandsamplelp}, the resulting number
of rows can be bounded by:
\begin{align*}
& O\left( \left(n^{\frac{1}{p} - \frac{1}{p'}}
        \tilde{n}^{\frac{1}{p'} - \frac{1}{2}} d^{\frac{1}{p}} R \right)^{p}
                d \log{d} \right)
 = O \left( n^{1 - \frac{p}{p'}}
        \left(d^{\frac{4}{p'}} \log^{\frac{2}{p'}}{d} \right)^{\frac{p}{p'} - \frac{p}{2}}
                R^{p} d^2 \log^{\frac{2 p}{p'} + 1}{d} \right)\\
\end{align*}
This leads to a result analogous to Lemma~\ref{lem:reductionlp}.
Solving for the fixed point of this process
allows us to prove Theorem~\ref{thm:rowsamplel1}.

\Proofof{Theorem~\ref{thm:rowsamplel1}}
Similar to the proof of Lemma~\ref{lem:rowsamplelp},
in each iteration we reduce the number of rows from $n$
to $O \left( n^{1 - \frac{p}{p'}}
        \left(d^{\frac{4}{p'}} \log^{\frac{2}{p'}}{d} \right)^{\frac{p}{p'} - \frac{p}{2}}
                R^{p} d^2 \log^{\frac{2 p}{p'} + 1}{d} \right)$.
Ignoring terms in $R$ and $\log{d}$, we have that the
number of rows converges doubly exponentially towards:
\begin{align*}
\left( \left(d^{\frac{4}{p'}}\right)^{\frac{p}{p'} - \frac{p}{2}}
                d^2\right)^{\frac{p'}{p}}
& = d^{\frac{4}{p'} - 2 + 2 \frac{p'}{p}}
= d^{\frac{4}{p'} + 2 p' - 2}
\end{align*}
This is minimized when $p' = \sqrt{2}$, giving
$\frac{4}{p'} + 2 p' - 2 = 4 \sqrt{2} - 2$.
As we can set $R = d^{O(\theta)}$,
the number of rows in $\matb$ can be bounded by
$O(d^{4 \sqrt{2} - 2 + \theta})$ for any constant $\theta$.
\QED

This method can be used to reduce the number of rows
for all values of $1 \leq p \leq 2$.
A calculation similar to the above proof leads to a row
count of $O(d^{\sqrt{\frac{8}{p}} - 2})$.
However, using three or more steps does not lead to
a significantly better bound since we can only obtain
samples with about $d$ rows when $p = 2$.
For $p \ge 4$, multiple steps of this approach also allows
us to compute $\poly(d)$ sized samples for any value of $p$.
We omit this extension as it leads to a significantly higher
row count.

%% file: Paper.bbl
\newcommand{\etalchar}[1]{$^{#1}$}

%% file: leverageproofs.tex
\section{Properties and Estimation of Stretch and Leverage Scores}
\label{sec:leverageproofs}

\subsection{Properties of Generalized Stretch}

We now give proofs for estimating leverage scores and
row sampling that we stated in
Sections~\ref{sec:algo2}~and~\ref{sec:pnorm}.

\Proofof{Fact \ref{fact:leveragesum}}
\begin{align}
\sum_{i = 1}^n \str{\mata}{\veca_i}
& = \sum_{i = 1}^n  \veca_i (\mata^T \mata)^{\dag} \veca_i^T \nonumber\\
& = \sum_{i = 1}^n  \trace{(\mata^T \mata)^{\dag} \veca_i^T \veca_i} \nonumber\\
& = \trace{(\mata^T \mata)^{\dag } \sum_{i = 1}^n \veca_i^T \veca_i} \nonumber \\
& = \trace{(\mata^T \mata)^{\dag } \mata^T \mata} = r
\end{align}
\QED

\Proofof{Fact \ref{fact:stretchclosedform}}
Note that both stretch and the Frobenius norm acts
on the rows independently.
Therefore it suffices to prove this when $\mata'$ has a single row,
aka. $\mata' = \veca$.
In this case the cyclic property of trace gives:
\begin{align}
\str{\mata}{\veca}
& = \veca (\mata^T \mata)^{\dag } \veca^T \nonumber \\
& = \veca (\mata^T \mata)^{\dag 1/2} (\mata^T \mata)^{\dag 1/2} \veca^T \nonumber \\
& = \nbr{\mata^{\dag 1/2} \veca^T}_2^2
\end{align}
\QED

\Proofof{Lemma \ref{lem:referenceswitch}}
The condition given implies that the null spaces
of $\matb_1^T \matb_1$ and $\matb_2^T \matb_2$ are identical, giving:
\begin{align}
(\matb_1^T \matb_1)^{\dag}
\preceq  (\matb_2^T \matb_2)^{\dag}
\preceq \kappa (\matb_1^T \matb_1)^{\dag}
\end{align}
Applying this to the vector $\vecx$ gives:
\begin{align}
 \vecx (\matb_1^T \matb_1)^{\dag} \vecx^T
\leq &  \vecx (\matb_2^T \matb_2)^{\dag} \vecx^T
\leq \kappa \vecx (\matb_1^T \matb_1)^{\dag} \vecx^T
\end{align}
\QED

\subsection{Estimation of Generalized Stretch}

Based on this fact, we can estimate these scores using
randomized projections in a way that's by now standard
\cite{SpielmanS08,DrineasMMW11}.
Pseudocode of our estimation algorithm is shown in Algorithm
\ref{alg:approxstretch},
while the error analysis is nearly identical to the ones given in
Section 4 of \cite{SpielmanS08} and Section 3.2. of \cite{DrineasMMW11}.

\begin{algo}[ht]
\qquad

$\textsc{ApproxStr}(\mata, \matb, \kappa, \projerror)$
\vspace{0.05cm}

\underline{Input:}
A $n\times d$ matrix $\mata$, approximation $m \times d$ matrix $\matb$ such that
$\frac{1}{\kappa} \mata^T \mata \preceq \matb^T \matb \preceq \mata^T \mata$,
parameter $\projerror \ge e^2$ indicating allowed estimation error.

\underline{Output:}
Upper bounds for stretches of rows of $\mata$
measured 
$\approxleverage_1 \ldots \approxleverage_n$.

\begin{algorithmic}
\STATE{Compute $\matc = (\matb^T \matb)^\invsqr$}
\STATE{Let $k = O(\log_{\projerror} (1 / \delta))$}
\STATE{Let $\matu$ be a $k \times d$ matrix with
        each entry is picked independently from $\Ncal(0, 1)$}
\STATE{Let $\approxleverage_i = \frac{\projerror}{d} \nbr{\matu \matc \veca_i^T}_2^2$.}
\RETURN{$\approxleveragev$}
\end{algorithmic}

\caption{Algorithm for Upper Bounding Stretch}

\label{alg:approxstretch}
\end{algo}

We remark that $(\matb^T \matb)^\invsqr$
can be replaced by any matrix whose product with its
transpose equals to $\matb^T \matb$.
nee candidate for this is $\matb (\matb^T \matb)^{\dag}$, and
using it would avoid computing the $1/2$ power of a matrix.
However, from a theoretical point of view both of these operations
take $O(m d^{\omega - 1})$ time, and we omit this extra step
for simplicity.



\Proofof{Lemma~\ref{lem:approxstretch}}

 Since $\matb^T\matb \preceq \mata^T\mata$ by assumption,
then $\rbr{\mata^T\mata}^\dag \preceq \rbr{\matb^T\matb}^\dag$. Denote by $\tau'_i =
\str{\matb}{\veca_i} = \veca_i \rbr{\matb^T \matb}^\dag \veca_i^T$,    note that $\tau_i =
\str{\mata}{\veca_i} = \veca_i \rbr{\mata^T \mata}^\dag \veca_i^T$, we have $\tau'_i \ge
\tau_i$. Next we show u $\tilde \tau_i \ge \tau'_i$  holds with large probability. 
By Lemma \ref{lem:jl} Part \ref{part:lower} we have:
\begin{align*}
  \Pr \sbr{\tilde \tau_i \le \tau'_i} &=  \Pr \left[\frac{1}{k} \nbr{\matu  \matc \veca_i^T}_2^2
        \le \frac{1}{\projerror } \nbr{  \matc \veca_i^T}_2^2\right]  \\
&\leq  \exp\left(\frac{k}{2} (1 - \projerror^{-1} -  \ln{\projerror})) \right) \\
&\leq  \exp\left(-\frac{k}{2} \frac{\ln{\projerror}}{2} \right) \\
& = \projerror^{-\frac{k}{4}},
\end{align*}
where the last inequality is due to the assumption that $R \le e^2$. 
By a suitable choice of constants in $k = O(\log_{\projerror}{nd^{c}})=O(\log_{\projerror}d)$
this can be made the above probability less than $n^{-1}d^{-c}$, taking a union bound over the $n$
rows gives Part~\ref{part:l2stretchupper}.

The upper bound on $\|\approxleveragev\|_1$ can be obtained
similarly. From Lemma~\ref{lem:referenceswitch}, we have $\tau'_i \le \kappa \tau_i$ holds
for all $i$. Then using Part \ref{part:upper} of Lemma \ref{lem:jl}:

\begin{align*}
  \Pr \sbr{\tilde \tau_i \ge R^2 \tau'_i} &= \Pr \sbr{ \frac{1}{k} \nbr{\matu \matc \veca_i^T}_2^2
    \ge {\projerror} \nbr{\matc \veca_i^T}_2^2} \\
&\leq  \exp\left(\frac{k}{2} (1 - \projerror +  \ln{\projerror}) \right) \\
&\leq  \exp\left(-\frac{k}{2} \ln{\projerror} \right) \\
& = \projerror^{-\frac{k}{2}},
\end{align*}
the last inequality is due to the fact that $R - 2\ln R$ increases w.r.t.~$R$ and $R-2\ln
R \ge 1$ holds when $R=e^2$. The above probability will be less than $n^{-1}d^{-c}$ if
choosing the same constants $k$ as before. Then $\| \approxleveragev\|_1 \le R^2 \| \leveragev' \|_1$
holds with probability at least $1-d^{-c}$. Together with $\| \leveragev' \|_1 \le \kappa \|
\leveragev \|_1$, then we obtain the upper bound of $\| \approxleveragev\|_1$.

It remains to bound the running time of $\textsc{LeverageUpper}$.
Finding $\matb^T\matb$ takes $O(n_b d^{\omega - 1})$ time,
while inverting it takes an additional $d^{\omega}$ time.
Computing $\matu \matc$ can be done in $O(kd^2)$ time,
and it It remains to evaluate $\matu \matc \veca_i^T$
for all rows $i$.
This can be done by summing $\nnz(\veca_i)$ length $k$ vectors,
giving a total of $\nnz(\mata) k$ over all $n$ vectors.
Therefore the total cost for computing the estimates is
$O(k(\nnz(\mata) + d^2) + (n_b + d)d^{\omega - 1})$.
\QED

\subsection{Estimation of $p$-Norm Leverage Scores}

We will estimate the values of $\nbr{\matu_{i*}}_p$ using similar
dimensionality reduction theorems.
Specifically, we utilize a result on $p$-stable distributions first
shown by Indyk \cite{Indyk06}.

\begin{lemma}
\label{lem:pstable}
(Theorem 4 of \cite{Indyk06})
For any $p \in (0, 2)$, any $c$ and any error factor $R$, there exist a
$d \times O(\log_{R} d)$ matrix $\Pi$ such that for any vector
$\vecz \in \Re^{d}$, we can obtain estimates $\approxleverage_i$
such that with probability $1 - d^{-c}$:
\begin{align*}
\frac{1}{R} \nbr{\vecz}_p \leq \approxleverage_i \leq R \approxleverage_i
\end{align*}
\end{lemma}

Note that the result from \cite{Indyk06} was only stated in terms of obtaining
$1 \pm \epsilon$ approximations, which leads to a factor of $\log{d}$
on the leading term.
However, these bounds can be obtained analogously by the fact
that $p$-stable distributions have bounded derivative.

We can now apply this projection matrix to $\mata \matc$ and
examine the rows of $\mata \matc \Pi^T$, which can in turn be
computed in $O(\nnz(\mata) \log_{R}d)$ time.
We can no longer use these projections when $p \geq 2$.
As a result, we will instead use the $L_2$ norm as an estimate
and apply the random projection given in Lemma~\ref{lem:jl}.
Fact~\ref{fact:holdersimple} gives that this leads to an extra
distortion by a factor of $O(d^{(|\frac{1}{2} - \frac{1}{p}|)p}) = O(d^{\frac{p}{2} - 1})$.
This distortion can be accounted for in the number of rows returned.
Pseudocode of this estimation and sampling routine is given in
Algorithm~\ref{alg:estimatesamplelp}

\begin{algo}[ht]
\qquad

$\textsc{EstimateAndSampleP}(\mata, \matc, \alpha, \beta, p, \epsilon)$
\vspace{0.05cm}

\underline{Input:}
$n \times d$ matrix $\mata$,
$\matc$ such that $\mata \matc$ is a $(\alpha, \beta, p)$
well-conditioned basis for $\mata$
projection error $R$, and output error $\epsilon$

\underline{Output:}
Matrix $\matb$

\begin{algorithmic}[1]
  \STATE{Compute projection matrix $\Pi \in \Re^{O(\log_{R}{d}) \times d}$}
  \STATE{Compute $d \times d_1$ matrix $\matc \Pi^T$}
  \STATE{Compute estimates of $\approxleveragep_i$
                from the rows of $\mata (\matc \Pi^T)$}
  \STATE{Compute probabilities $p_i = O\left(R^{2p} (\alpha \beta)^{p} \frac{\approxleveragep_i}{\sum_{i} \approxleveragep_i}\right)$}
  \RETURN{$\textsc{Sample}(\mata, p, \epsilon)$}
\end{algorithmic}

\caption{Leverage Estimation and Sampling Routine for $p$-norm}

\label{alg:estimatesamplelp}
\end{algo}

\Proofof{Lemma~\ref{lem:estimateandsamplelp}}
Applying a union bound over the $n = \poly(d)$ rows of $\mata$
gives that with probability at least $1 - d^{-c - 1}$, we have:
\begin{align*}
\nbr{\mata \matc_{i*}}_p^p \leq R^{p} \approxleveragep_i\\
\approxleveragep_i \leq R^{p} \nbr{\mata \matc_{i*}}_p^p
\end{align*}
Therefore we have:
\begin{align*}
 R^{2p} \frac{\approxleveragep_i}{\sum_{i} \approxleveragep_i}
\geq & \frac{\nbr{\mata \matc}_p^p}{\sum_{i} \approxleverage_i}
\geq \frac{\nbr{\mata \matc}_p^p}{\sum_{i}  \nbr{\mata \matc_{i*}}_p^p}
\end{align*}
The guarantees on the output then follows from computing
Lemma~\ref{lem:samplelp}, while the total running time follows
from the cost of evaluating $\mata (\matc \Pi^T)$.
\QED

%% file: rowcombineproofs.tex
\section{Deferred Proofs from Section~\ref{sec:algo2}}
\label{sec:rowcombineproofs}

\Proofof{Lemma \ref{lem:goodapprox}}
Let $\matc = \rbr{\mata^T \mata}^\invsqr$,
by Fact \ref{fact:stretchclosedform} we have that:
$
\str{\mata}{\mata_{(b)}}
= \|\mata_{(b)} \matc\|_F^2
$
and
$
\str{\mata}{\matashort_{(b)}}
= \|\matashort_{(b)} \matc\|_F^2
$.
Furthermore, since $\matashort_{(b)} = \matu_{(b)} \mata_{(b)}$,
we have:
\begin{align*}
\str{\mata}{\matashort_{(b)}}
= \|\matu_{(b)} \mata_{(b)} \matc\|_F^2
\end{align*}

Next we upper bound this term by using similar technology from the proof of
Lemma~\ref{lem:approxstretch}. Let $\vecy_i$ be the $i$-th column of
$\mata_{(b)}\matc$. As the entries of $\matu_{(b)}$ are independent standard Gaussian
random variables, 
it is easy to see that $\mathbb{E} \sbr{ \nbr{\matu_{(b)} \vecy_i}_2^2} =
k\nbr{\vecy_i}_2^2$ for any $i$. 
By Lemma \ref{lem:jl}, we have that:
\begin{align*}
 \prob{\|\matu_{(b)} \vecy_{i} \|_2^2 
         \le \frac{k}{ R} \|\vecy_{i}\|_2^2}
       \leq  \exp \left( \frac{k}{2} \rbr{1-R^{-1} - \ln{R}} \right)
       \le R^{-\frac{k}{4}}
\end{align*}
Where the last inequality is by $1-R^{-1} - \ln{R}\le -\frac{1}{2}\ln{R}$ with
the assumption that $R\ge e^2$.
If we $k$ to $4(c+1)\theta^{-1}\log_d n$ and substitute $R = d^\theta$,
we get:
\begin{align*}
  R^{{-\frac{k}{4}}} \le d^{-\frac{k\theta}{4}} = d^{-c-1}n^{-1}.
\end{align*}
Using the union bound, we have
\begin{align*}
  \Pr \sbr{\str{\mata}{\matashort_{(b)}} \le \frac{k}{R} \str{\mata}{\mata_{(b)}}} =
  \Pr \sbr{\sum_{i=1}^d \|\matu_{(b)} \vecy_{i} \|_2^2  \le \sum_{i=1}^d \frac{k}{ R}
    \|\vecy_{i}\|_2^2}  
\le d^{-c}n^{-1}
\end{align*}
Apply the union bound again, the above holds for all $b=1,\ldots,n_b$ with probability at
most $d^{-c}$.
By the assumption that $n=\poly(d)$, $k = O(c/\theta)$ suffices.
\QED

\Proofof{Lemma \ref{lem:shortupper}}
By definition we have $\matashort_{(b)} = \matu_{(b)} \mata_{(b)}$.
Let $\vecx$ be any vector in $\Re^{d}$, we first have 
\begin{align*}
\|\matu_{(b)} \vecx\|_2^2 
\leq \|\matu_{(b)}\|_2^2  \|\vecx\|_2^2  \le \|\matu_{(b)}\|_F^2  \|\vecx\|_2^2 
\end{align*}
and then
\begin{align*}
 \vecx^T \matashort_{(b)}^T \matashort_{(b)} \vecx
 & = \|\matashort_{(b)} \vecx\|_2^2 \\
 & = \|\matu_{(b)}\mata_{(b)} \vecx\|_2^2 \\
& \leq   \|\matu_{(b)}\|_F^2  \|\mata_{(b)} \vecx\|_2^2\\
& =\|\matu_{(b)}\|_F^2 \cdot \vecx^T \mata_{(b)}^T\mata_{(b)} \vecx
\end{align*}
\QED


\Proofof{Lemma \ref{lem:pseudoinversereverse}}
Consider an orthonormal basis for the range space of $\matc$,
$\vecv_1 \ldots \vecv_{\textrm{rank}(\matc)}$.
Since $\matc + \matd \succeq \matc$, this basis
can be extended to an orthonormal basis to the range space
of $\matc + \matd$ by adding
$\vecv_{\textrm{rank}(\matc)+1} \ldots \vecv_{\textrm{rank}(\matc + \matd)}$.
It suffices to prove the claim under this basis system.
Here $\matc$ and $\matd$ can be rewritten as by proper rotation:
\begin{align*}
\matc = 
\begin{bmatrix}
\matc_{11} & \matzero \\
\matzero & \matzero
\end{bmatrix}
\textrm{ and }
\matd = 
\begin{bmatrix}
\matd_{11} &  \matd_{12}  \\
\matd_{12}^T &  \matd_{22}  \\
\end{bmatrix},
\end{align*}
where $\matc_{11}$ and $\matd_{22}$ are strictly
positive definite.
Furthermore, since $\matd$ is positive semi-definite we have that
$\matd_{11} - \matd_{12} \matd_{22}^{-1} \matd_{12}^T$ is
also positive semi-definite.
For any vector $\vecx$, $\matproj_\matc \vecx$ gives a vector that's non-zero
only in the first $\textrm{rank}(\matc)$ entries.
Let this part be $\vecx_{1}$.
Then evaluating $(\matc + \matd)^{\dag} \matproj_\matc \vecx =
[ \vecy_{1} ; \vecy_{2} ]$
becomes solving the following system:
\begin{align*}
(\matc_{11} + \matd_{11}) \vecy_1 + \matd_{12}  \vecy_2 &=  \vecx_1 \\
\matd_{12}^T \vecy_1  + \matd_{22} \vecy_2 &=  \veczero
\end{align*}
The second equation gives $\vecy_2 =
-\matd_{22}^{-1} \matd_{12}^T \vecy_1$.
Substituting it into the first one gives:
\begin{align*}
\rbr{\matc_{11} + \matd_{11}
- \matd_{12} \matd_{22}^{-1} \matd_{12}^T} \vecy_1 = \vecx_1
\end{align*}
Note that this is the same as taking the partial Cholesky
factorization onto the range space of $\matproj_\matc$.
Combining things gives:
\begin{align*}
\vecx^T \matproj_\matc (\matc + \matd)^{\dag} \matproj_\matc \vecx
=  \vecx_{1}^T (\matc_{11} + \matd_{11}
- \matd_{12} \matd_{22}^{-1} \matd_{12}^T  )^{-1}
\vecx_{1}
\end{align*}
Since both $\matd_{11} - \matd_{12} \matd_{22}^{-1} \matd_{12}^T$ and $\matd_{11}$
are positive definite, we have
$\matc_{11} \preceq \matc + \matd_{11} - \matd_{12} \matd_{22}^{-1} \matd_{12}^T$ and therefore:
\begin{align*}
\vecx^T \matproj_\matc (\matc + \matd)_2^\dag \matproj_\matc \vecx
&=  \vecx_1^T (\matc_{11} + \matd_{11}
- \matd_{12} \matd_{22}^{-1} \matd_{12}^T)
  \vecx_1^T\nonumber \\
&\le  \vecx_1^T \matc_{11}^{-1} \vecx_1^T \nonumber \\
&=  \vecx^T \matproj_\matc \matc^{+} \matproj_\matc \vecx
\end{align*}
holds for every $\vecx$. 
\QED